\documentclass[graphics,floatfix, footinbib,tightenlines,nobibnotes, aps, prb, twocolumn]{revtex4-2}

\usepackage{amsmath}
\usepackage{amssymb}
\usepackage{graphicx}% Include figure files
\usepackage{dcolumn}% Align table columns on decimal point
\usepackage{comment}
\usepackage{bm}% bold math
\usepackage{braket}
\usepackage{subcaption}
\usepackage{verbatim}
\usepackage{float}
\usepackage{bbold}
\usepackage{color}
\usepackage{xcolor}
\usepackage{relsize}
\usepackage{amsthm}
\usepackage{enumerate}
\usepackage{soul,xcolor}
\usepackage[T1]{fontenc}
\usepackage[colorlinks=true ,urlcolor=blue,urlbordercolor={0 1 1}]{hyperref}

\newcommand{\beq}{\begin{eqnarray}}
\newcommand{\eeq}{\end{eqnarray}}

\newcommand{\bsp}{\begin{split}}
\newcommand{\esp}{\end{split}}

\newcommand{\be}{\begin{equation}}
\newcommand{\ee}{\end{equation}}

\begin{document}
\title{Quasiperiodicity, band topology, and moir\'e graphene}
\author{Dan Mao}
\affiliation{Department of Physics, Massachusetts Institute of Technology, Cambridge Massachusetts
02139, USA.}
\author{T. Senthil}
\affiliation{Department of Physics, Massachusetts Institute of Technology, Cambridge Massachusetts
02139, USA.}

\begin{abstract}
   A number of moir\'e graphene systems have nearly flat topological bands where electron motion is strongly correlated. Though microscopically these systems are only quasiperiodic,  they can typically be treated as translation invariant to an excellent approximation. Here we reconsider this question for magic angle twisted bilayer graphene that is nearly aligned with a hexagonal boron nitride(h-BN) substrate. We carefully study  the effect of the periodic potential induced by h-BN on the low energy physics. The combination of this potential and the moir\'e lattice produced by the twisted graphene  generates a quasi-periodic term  that depends on the alignment angle between h-BN and the moir\'e graphene. We find that the alignment angle has a significant impact on both the band gap near charge neutrality and the behavior of electrical transport. We also introduce  and study toy models to illustrate how a quasi-periodic potential can give rise to localization and change in  transport properties of topological bands.
\end{abstract}
\maketitle
\section{Introduction}
Following the discovery of correlated insulators and superconductivity in Magic Angle Twisted Bilayer Graphene (MATBG) in 2018\cite{cao2018unconventional,cao2018correlated} a tremendous amount of attention has been lavished on moire materials. Other moire systems displaying correlated electron physics  include ABC Trilayer Graphene (TLG/hBN) nearly aligned with a hexagonal Boron-Nitride (hBN) substrate\cite{chen2020tunable}, twisted double bilayer graphene\cite{shen_correlated_2020}, twisted monolayer-bilayer graphene\cite{chen2020electrically}, and twisted transition metal dichalcogenides\cite{zhang_flat_2020}. Our interest in this paper is on  MATBG that is further nearly aligned with a hBN substrate (MATBG/hBN)\cite{sharpe_emergent_2019,serlin_intrinsic_2020} which alters the observed phenomena. 

In MATBG/hBN Ref.\onlinecite{sharpe_emergent_2019} discovered ferromagnetism and an associated large anomalous Hall effect  at $3/4$ filling of the conduction band. Subsequently Ref.\onlinecite{serlin_intrinsic_2020} studied devices of MATBG/hBN which not only showed emergent ferromagnetism at $3/4$ conduction band filling but also observed a quantized anomalous Hall effect with $\sigma_{xy} = \frac{e^2}{h}$. Theoretically the near alignment with the hBN breaks the $C_2$ symmetry of 180 degree rotation within the graphene plane and opens up a gap between the valence and conduction bands which - in the absebnce of alignment - touch at Dirac points. The resulting bands within a single valley were found\cite{bultinck2020mechanism,zhang_twisted_2019} to have Chern number $\pm 1$ (with opposite valleys having opposite Chern number). As discussed in Ref. \onlinecite{zhang_nearly_2019} such nearly flat $\pm$ Chern bands are, in fact, common to a number of moire graphene materials. Upon including electron-electron interactions, Ref. \onlinecite{zhang_nearly_2019} also proposed these systems to be excellent platforms to show a quantum anomalous Hall effect at total ({\em i.e} including spin and valley) odd integer filling. These ideas were developed further in the specific context\cite{bultinck2020mechanism,zhang_twisted_2019} of MATBG/hBN, and in ABC TLG/hBN which too displays emergent ferromagnetism and a quantum anomalous Hall effect\cite{chen2020tunable}.

In this paper we revisit the theory of single particle states of MATBG/hBN. 
 The presence of h-BN layer has two effects on the nearby graphene. One is that h-BN induces a constant sub-lattice potential difference, which is studied in detail in \cite{bultinck2020mechanism,zhang_twisted_2019}. The other is that it induces a second periodic moir\'e potential which may or   may not be commensurate with the original moir\'e potential of the TBLG system. The previous theoretical work ignored the moire potential introduced by the near alignment with the hBN, mostly for simplicity but also on the grounds that its estimated strength is smaller than the TBLG moire potential. In the present paper we go beyond this approximation, and carefully include both moire potentials.  We first determine the conditions - which we dub ``perfect alignment" - under which the two moire potentials are commensurate. This concept of perfect alignment is distinct from  the naive expectation that the perfect situation is when the twist angle between one graphene layer and hBN is zero. When the perfect alignment condition is satisfied , 
translation invariance is preserved and we can define a crystal momentum and a (reduced) Brillouin zone. Away from perfect alignment, the two moire potentials are  incommensurate case, and  translational symmetry is completely broken. The low energy physics can be modeled by introducing a quasi-periodic potential to  topological bands (in the case of TBLG/hBN, Chern bands with opposite Chern numbers). 

Electronic systems with a quasi-periodic potential(QP) has been studied extensively in 1D. (See \cite{sokoloff_unusual_1985} for a detailed review.) In the 1D Audry-Andr\'e model, there is a localization transition with the increase of quasi-periodic potential strength.\cite{aubry1980analyticity} In higher than 1D, an intermediate phase with eigenstates delocalized in both real space and momentum space can exist between an extended phase and a localized phase.\cite{devakul_anderson_2017} The generic existence of such an intermediate phase in a 2D system with quasi-periodic potential has not been settled yet but it is not our focus in this paper. We are particularly interested in the effect of a quasi-periodic potential on topological bands\cite{fu2020flat}.  In momentum space, non-vanishing Chern number can impose non-trivial phase structure on the wave function, which may change the localization properties when a quasi-periodic potential is added to the system compared to trivial bands.

In the case of perfect alignment, there is a clean separation between valence and conduction bands. Then if - due to interactions - the system is valley and spin polarized at total odd integer filling $\nu_T$, electrons will completely fill a Chern band, and there will be a quantum anomalous Hall effect.  Away from perfect alignment, the quasiperiodic potential induces   in-gap states - which are not real-space localized - between the valence and conduction bands.  Then we show by explicit calculation  that even with full valley and spin polarization at odd integer $\nu_T$, there is no quantization of the anomalous Hall conductivity. Thus observation of a quantum anomalous Hall effect at such fillings is aided by studying devices that are tuned close to perfect alignment.  We show however that strain can be used to tune the alignment condition, thereby enabling engineering flat well separated Chern bands in TBLG/hBN devices. 

Though we do not address many body effects in this paper, we note that the presence of in-gap  states are likely to hinder the development of valley/spin polarization in the first place. This is because they can roughly be thought of as increasing the bandwidth of the active valence or conduction band, thereby reducing the ability of interactions to induce ferromagnetism. Thus it is desirable to stay close to perfect alignment. Indeed the two devices studied in Refs. \onlinecite{sharpe_emergent_2019,serlin_intrinsic_2020} are nearly perfectly aligned. This condition may be a more stringent requirement for the fractional quantum anomalous Hall states proposed\cite{repellin2020chern,ledwith2020fractional,abouelkomsan2020particle} for TBLG/hBN. 

 The periodic modulation induced by the hBN layer is relevant only if the hBN layer is nearly aligned with TBLG since the moir\'e lattice constants of the superlattices generated by hBN and  TBLG is of the same order as the moir\'e lattice constant of TBLG.
For hBN misaligned with TBLG, due to the lattice mismatch, there is no longer any periodic moir\'e potential induced by hBN so the QP physics are irrelevant in those systems.

 In recent years, Anderson localization and many-body localization in the presence of QP have been investigated in cold atom experiments\cite{roati2008anderson, deissler2010delocalization,schreiber2015observation, bordia2017probing}. The interplay between quasiperiodicity and interaction  near critical points in quantum Ising and related spin models have been the subject of several  studies: see, eg,  Refs. \onlinecite{luck1993classification,luck1993critical,igloi1988quantum,crowley2018quasiperiodic,crowley2019quantum,agrawal2020universality} for some representative papers. It is seen that the presence of QP can lead to new interacting critical phases which are different from that found with quenched disorder \cite{agrawal2020universality}. 
 The specific moir\'e graphene system we study here  provides an experimental context where strongly interacting quantum phases/phase transitions in the presence of quasiperiodicity may be explored.  

The rest of the paper is organized as follows. In section \ref{sec:geometry}, we explain how the alignment to hBN induces another moir\'e pattern on top of the original moir\'e pattern of TBLG system. We further study two scenarios in section \ref{sec:perfect} and in section \ref{sec:incomm}. One is that the two moir\'e patterns overlap and the other is that they are incommensurate. In section \ref{sec:toy}, we propose a toy model to address the question of the effect of a quasi-periodic potential on a topological band.

\section{Two moir\'e patterns in hBN/TBLG system}
\label{sec:geometry}
Let us consider TBLG with the top graphene layer nearly aligned with hBN. There are two moir\'e patterns, one formed by the TBLG, the other formed by top graphene layer and h-BN layer. The difference between the two moir\'e reciprocal lattice vectors is in general not small compared to the reciprocal vectors themselves. Thus, strictly speaking, it is not a valid approximation to define a mini-BZ. Let us first write down the reciprocal vectors explicitly. The reciprocal lattice vectors of the top graphene sheet are $\Vec{G}_{t,1} = \frac{4\pi}{\sqrt{3}a_G} (0,1)$ and $\Vec{G}_{t,2} = R_{2\pi/3}\Vec{G}_{t,1}$, where $a_G$ is the lattice constant of graphene and $R_{2\pi/3}$ denotes counter-clockwise rotation by $2\pi/3$.  Assuming the bottom graphene layer rotates counter-clockwise by an angle $\theta_G\ll1$, that gives, $\Vec{G}_{b,1} = R_{\theta_G} \Vec{G}_{t,1}$. The TBLG moir\'e pattern is determined by the two reciprocal lattice vectors, $\Vec{G}_{1} = \Vec{G}_{t,1}-\Vec{G}_{b,1} = \frac{4\pi}{\sqrt{3} a_G} (\sin{\theta_G},1-\cos{\theta_G})$ and $\Vec{G}_{2} = R_{2\pi/3}\Vec{G}_1$.
Now adding h-BN on top, assuming h-BN layer rotates by an angle $\theta_{BN}\ll 1$ with respect to the top layer of TBLG, there is a second moir\'e pattern, which is generated by the lattice mismatch of the h-BN layer and the top graphene layer. For the reciprocal lattice vectors for this second moir\'e pattern, we write,
$\Vec{Q}_{1} = \Vec{G}_{t,1}-\Vec{G}_{BN,1} = (\frac{4\pi}{\sqrt{3} a_{BN}}\sin{\theta_{BN}},\frac{4\pi}{\sqrt{3} a_{BN}}\cos{\theta_{BN}}-\frac{4\pi}{\sqrt{3} a_{G}})$, $\Vec{Q}_{2} = R_{2\pi/3} \Vec{Q}_1$, where $a_{BN}$ is the lattice constant of h-BN.

\begin{figure}[h]
\subfloat[]{
  \includegraphics[height = .2\textwidth]{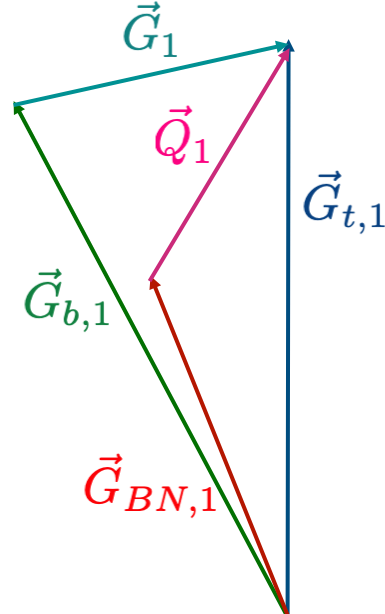}}
\subfloat[]{
  \includegraphics[height = .2\textwidth]{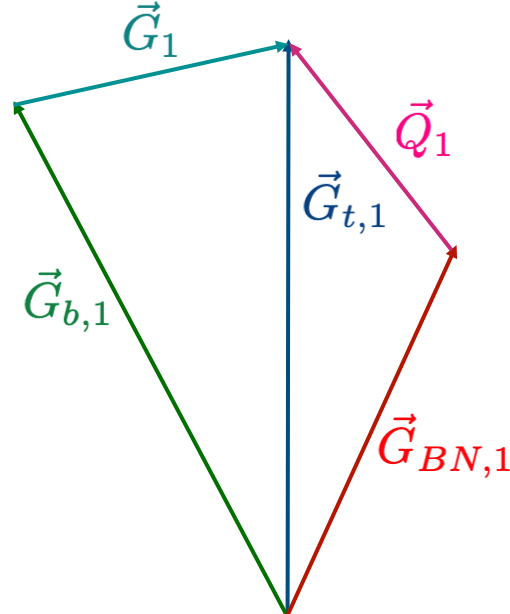}}
  \caption{(a)Case 1, $\theta_{BN}>0$ and $\theta_G>0$. (b)Case 2, $\theta_{BN}<0$ and $\theta_G>0$. The angles are exaggerated for illustration purpose.}
\label{fig:config}
\end{figure}

For special combination of $\theta_{BN}$ and $\theta_G$, the two moir\'e patterns can be commensurate. For simplicity, we only consider the case where these two patterns overlap, which we call "perfect" alignment. This means that the lattice generated by $\Vec{G}_{1,2}$ is the same as the lattice generated by $\Vec{Q}_{1,2}$, which can be satisfied as long as $|\Vec{G}_1| = |\Vec{Q}_1|$ and the angle between $\Vec{Q}_1$ and $\Vec{G}_1$ is $n\pi/3$, where $n$ is an integer. These two conditions can be satisfied either when $\theta_{G}$ and $\theta_{BN}$ have the same sign or have the opposite sign. See Fig.\ref{fig:config} for illustration.

\begin{figure}[h]
\subfloat[]{
  \includegraphics[width = .48\textwidth]{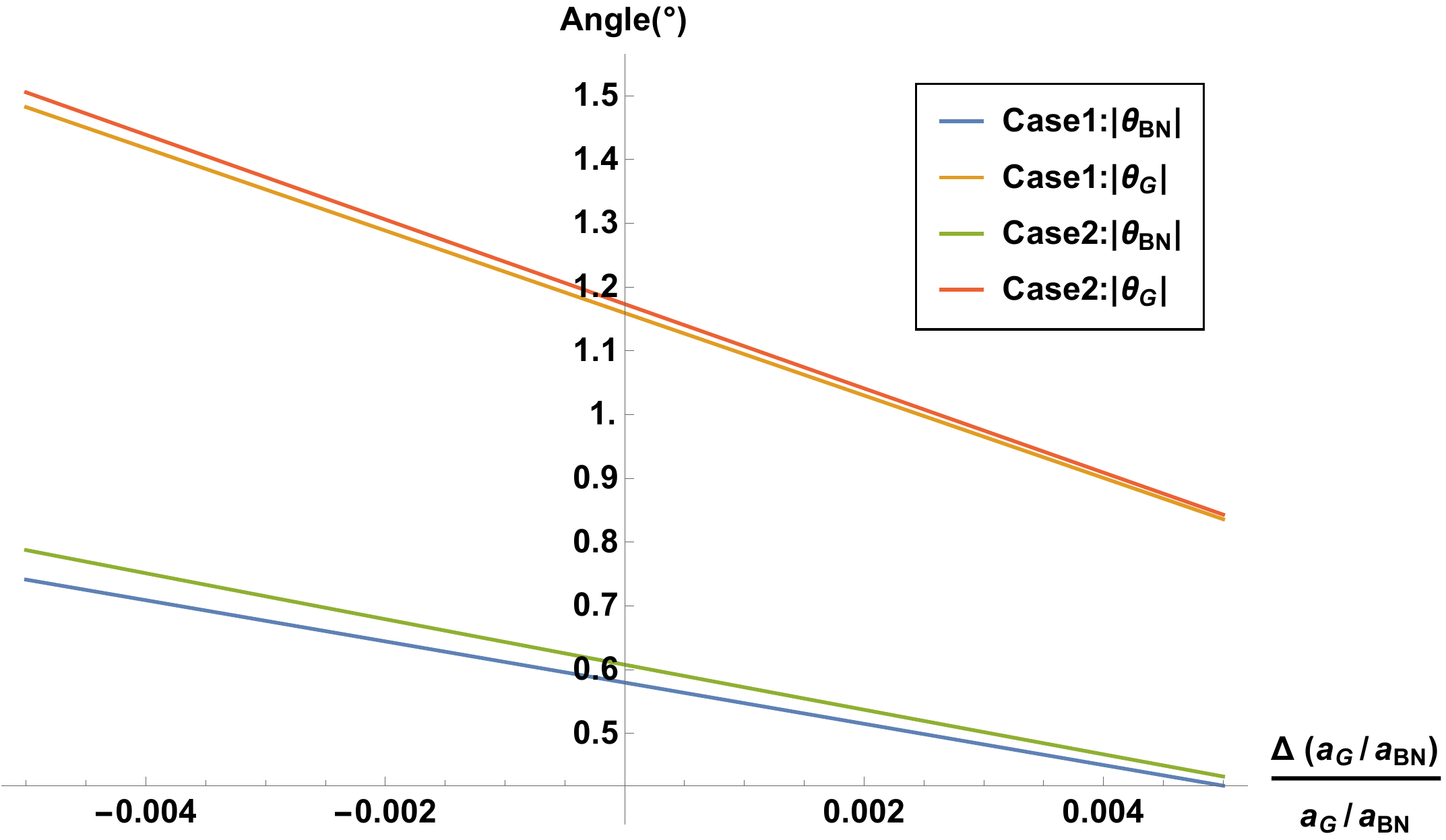}}
 
\subfloat[]{\includegraphics[width = .48\textwidth]{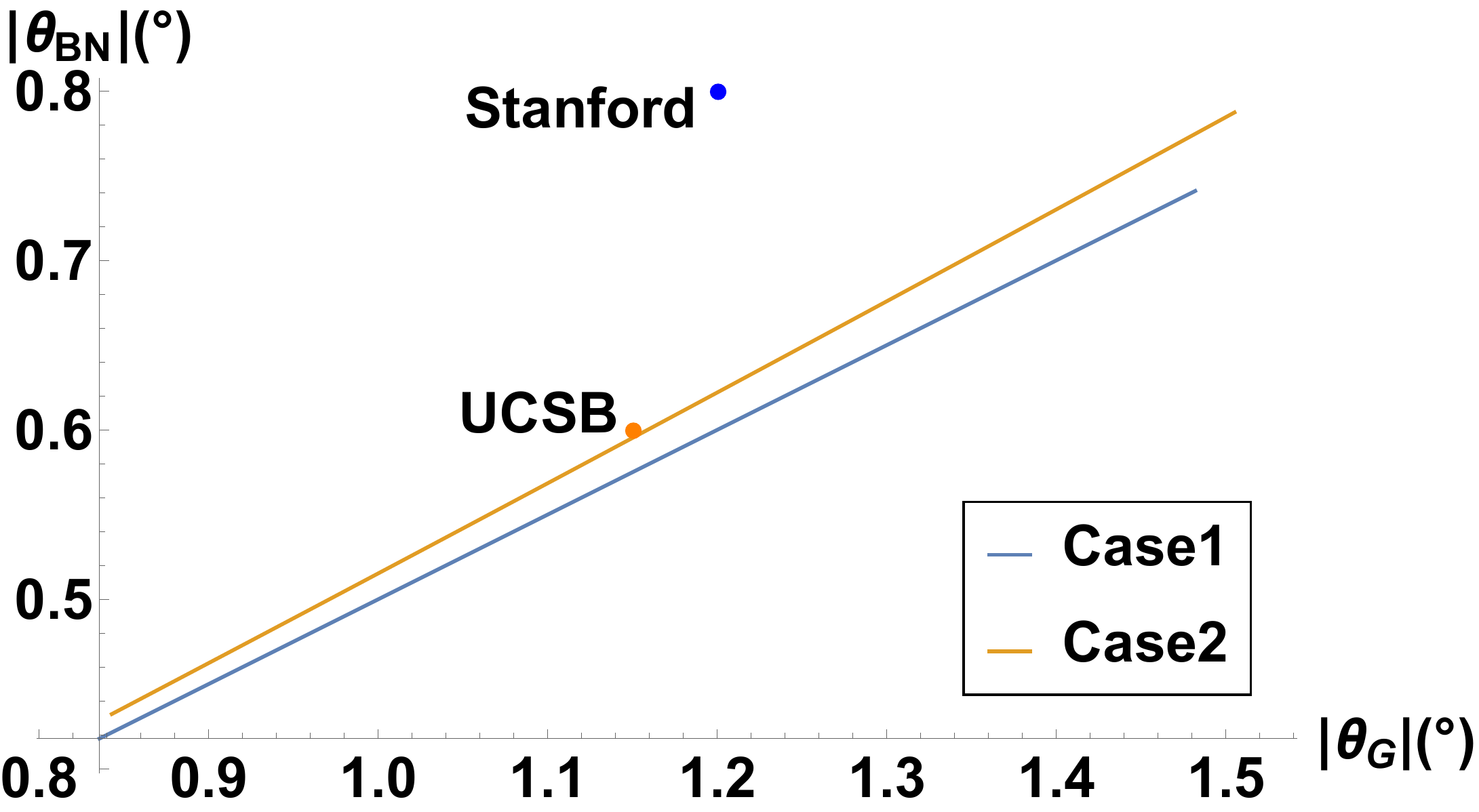}}
  \caption{(a)Dependence of $|\theta_{BN}|$ and $|\theta_{G}|$ on the change of the ratio $a_G/a_{BN}$ for satisfying perfect alignment conditions. $\Delta(a_G/a_{BN})$ denotes the change of $a_G/a_{BN}$ from $a_G/a_{BN} = 2.46/2.504$.
  (b)The points on blue and orange lines are $|\theta_{BN}|$ and $|\theta_{G}|$ taken from (a). The blue and orange dots are data from two experimental samples from the Stanford group \cite{sharpe_emergent_2019} and the UCSB group \cite{serlin_intrinsic_2020}. }
\label{fig:align}
\end{figure}

For case 1, $\theta_{BN}>0$ and $\theta_G>0$. The angle between $\Vec{G}_1$ and $\Vec{Q}_1$ is $\pi/3$. For case 2, $\theta_{BN}<0$ and $\theta_G>0$. The angle between $\Vec{G}_1$ and $\Vec{Q}_1$ is $2\pi/3$. We only consider $a_{BN} > a_G$. Using the law of sines one can get,
\begin{equation}
    \begin{cases}
        \cot{|\theta_{BN}|} = (2\sin{(\frac{\theta_G}{2})}\sin{(\mp\frac{\theta_G}{2}+\frac{\pi}{6}}))^{-1}-\cot{(\mp\frac{\theta_G}{2}+\frac{\pi}{6})}\\
        \frac{a_{G}}{a_{BN}} = \frac{2\sin{(\frac{\theta_G}{2})}\sin{(\mp\frac{\theta_G}{2}+\frac{\pi}{6}})}{\sin{|\theta_{BN}|}},
    \end{cases}
    \label{eq:align}
\end{equation}
where we take "$-$" for case 1 and "$+$" for case 2.

From the perfect alignment conditions Eq.\ref{eq:align}, there are two free parameters in $a_G$, $a_{BN}$, $\theta_{G}$ and $\theta_{BN}$. If one fix $a_{BN}$ to be $2.504$\AA~and $a_{G}$ to be $2.46$\AA, the corresponding $\theta_{G,BN}$ are $\theta_{G} \approx 1.16^\circ$, $\theta_{BN}\approx 0.58^\circ$ for case 1 and $\theta_{G} \approx 1.17^\circ$, $\theta_{BN}\approx -0.61^\circ$ for case 2. If the graphene sheets are under strain, $a_G$ can be slightly changed. From Fig.\ref{fig:align}, we find that $\theta_{BN}$ and $\theta_{G}$ are highly sensitive to the lattice constants and $\theta_G$ can be tuned to magic angle with $0.2\%$ change to $a_G/a_{BN}$. We also plot the experimental value of $\theta_{G}$ and $\theta_{BN}$\cite{sharpe_emergent_2019,serlin_intrinsic_2020} in Fig.\ref{fig:align}(b) and compare them to the perfect alignment case. Note that the sample studied in Ref.\cite{serlin_intrinsic_2020} is closer to the perfect alignment. This provides an explanation to the better quantization of Hall conductivity in Ref.\cite{serlin_intrinsic_2020} than in Ref.\cite{sharpe_emergent_2019}.

\begin{comment}
\begin{table}[]
\centering
\begin{tabular}{|c|c|c|c|c|}
\hline
$\theta_G$ &\multicolumn{2}{c|}{$1.2^\circ$} & \multicolumn{2}{c|}{$1.15^\circ$ } \\
\hline
 & Case 1 & Case 2 & Case 1 & Case 2 \\
\hline
$\theta_{BN}$&$0.60^\circ$ & $-0.62^\circ$ & $0.57^\circ$ & $-0.60^\circ$\\
\hline
$a_{BN}/$\AA & 2.506 & 2.505 & 2.504 & 2.503\\
\hline
\end{tabular}
\caption{Perfect alignment condition for two different $\theta_{G}$'s, where the lattice constant of graphene is taken to be $a_{G} = 2.46$\AA.}
\label{tab:alignment_cond}
\end{table}
In experiments, since the lattice constant of $a_{BN}$ is fixed, in order to satisfy the perfect alignment condition, one needs to tune both $\theta_{BN}$ and $\theta_{G}$. The observed value for $a_{BN} \approx2.504$\AA, compared to Table.\ref{tab:alignment_cond}, is within 0.1\% in the perfect alignment condition range for $\theta_G = 1.2^{\circ}$ and $1.15^{\circ}$ in various experiments\cite{sharpe_emergent_2019,serlin_intrinsic_2020} and $\theta_{BN}$'s can be around $0.8^{\circ}$ or $0.6^{\circ}$ in these experiments.
\end{comment}

Next, we continue to discuss the two kinds of perfect alignments in details in section \ref{sec:perfect}.

\section{Perfect alignment}
\label{sec:perfect}
As an idealized limit, in this section, we consider perfect alignment between h-BN and TBLG. We can still define mini Brillouin zone and momentum is a good quantum number in this limit. We ignore the hopping between h-BN layer and the bottom layer of the TBLG system. Hopping between h-BN and top layer of TBLG induces two kinds of terms in momentum space of the graphene. One is hopping terms between $\vec{k}$ and $\vec{k}+\vec{Q}_i$'s, where $\vec{Q}_i$'s are the reciprocal vectors of the moir\'e pattern generated by h-BN and top graphene layer. The other one is a constant AB sublattice potential due to the lattice relaxation in h-BN and in graphene and electron-electron interaction (If the lattice is rigid, the sublattice potential vanishes due to the lattice mismatch between h-BN and graphene.).\cite{jung_origin_2015} In momentum space, the Hamiltonian of the h-BN and TBLG system for one valley and one spin can be written as,
\begin{equation}
    H = H_{TBLG} + H_V,
\end{equation}
where $H_V$ contains two terms,
\begin{equation}
    H_V = \sum_{\vec{k}} f_{\vec{k}}^\dag m_z \sigma_z f_{\vec{k}} + \sum_{\vec{k},i} (f_{\vec{k}}^\dag V(\vec{Q}_i) f_{\vec{k}+\vec{Q}_i}+h.c.),
\end{equation}
where $f_{\vec{k}} = (f_{\vec{k},A}, f_{\vec{k},B})^T$ denotes the electron annihilation operators for sublattice A and B. $\sigma_z$ acts on sublattice degree of freedom. Index $i = 1,...,6$ labels different reciprocal vectors. $\vec{Q}_1$ is defined in section \ref{sec:geometry} and all the other $\vec{Q}_i$'s are generated by performing $\mathcal{C}_6$ rotation of $\vec{Q}_1$ consecutively.

$V(\vec{Q}_i)$'s can be parametrized in the following way\cite{jung_ab_2014},
\begin{equation}
V(\vec{Q}_i)= \left(
    \begin{array}{cc}
       H_0(\vec{Q}_i) + H_z(\vec{Q}_i) & H_{AB}(\vec{Q}_i)\\
        H_{BA}(\vec{Q}_i) &H_0(\vec{Q}_i)-H_z(\vec{Q}_i)
    \end{array}
    \right),
    \label{eq:qp}
\end{equation}
where $H_{0,z}(\vec{Q}_1) = H_{0,z}(\vec{Q}_3)=H_{0,z}(\vec{Q}_5) = C_{0,z} e^{i\phi_{0,z}}$, $H_{0,z}(\vec{Q}_2) = H_{0,z}(\vec{Q}_4)=H_{0,z}(\vec{Q}_6) = C_{0,z} e^{-i\phi_{0,z}}$ and $H_{AB}(\vec{Q}_1) = H_{AB}^*(\vec{Q}_4) = C_{AB} e^{i(\frac{2\pi}{3}-\phi_{AB})}$, $H_{AB}(\vec{Q}_3) = H_{AB}^*(\vec{Q}_2) = C_{AB} e^{-i\phi_{AB}}$, $H_{AB}(\vec{Q}_5) = H_{AB}^*(\vec{Q}_6) = C_{AB} e^{i(-\frac{2\pi}{3}-\phi_{AB})}$. $H_{BA}(\vec{Q}) = H_{AB}^*(-\vec{Q})$ from Hermiticity. 

From ab initio study\cite{jung_ab_2014}, at $\theta_{BN} = 0^\circ$, taking the lattice relaxation into account, the parameters for the periodic terms are
$C_0 = -9.07$, $\phi_0 = 97.99^\circ$, $C_z = -5.64$,$\phi_z = -3.66^\circ$, $C_{AB} = 7.34$, $\phi_{AB} = 24.53^\circ$. All $C$'s are in units of $meV$.  We take $m_z=15$ meV in the numerics. In our cases, $\theta_{BN}$ is not always zero, but we adopt the above set of parameters, assuming that the slight change will not alter the low energy physics.

Compared to the pure TBLG system, the alignment of hBN layer can in principle  open up a gap at $K_M$ points in the mini BZ due to the breaking of $C_2\mathcal{T}$ symmetry induced by h-BN.\cite{Po_origin,Zou_band,Po_fragile}

\begin{figure}[h]
\subfloat[]{
  \includegraphics[width = .4\textwidth]{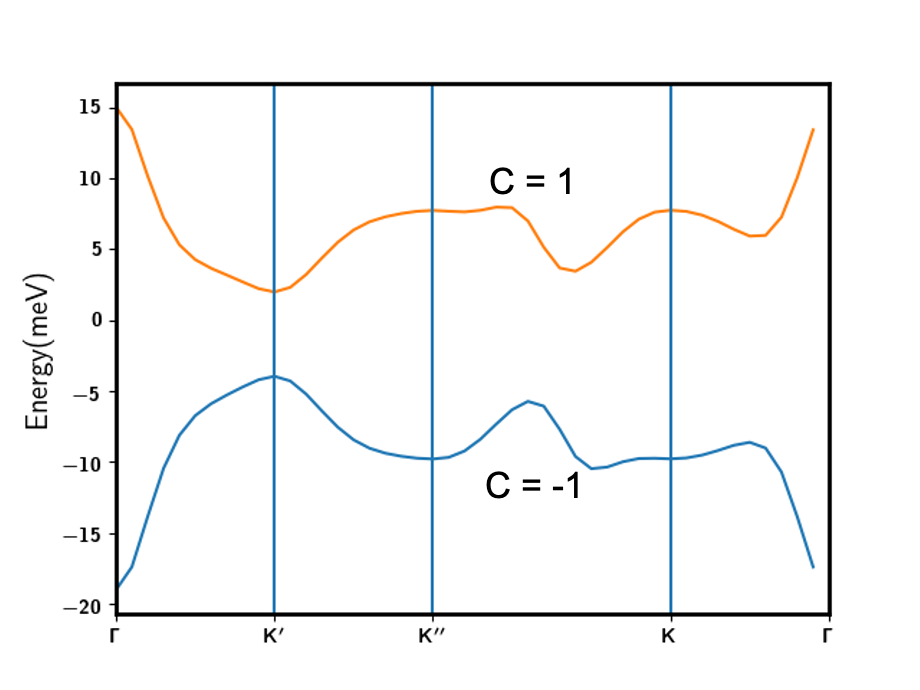}}
  
\subfloat[]{
  \includegraphics[width = .4\textwidth]{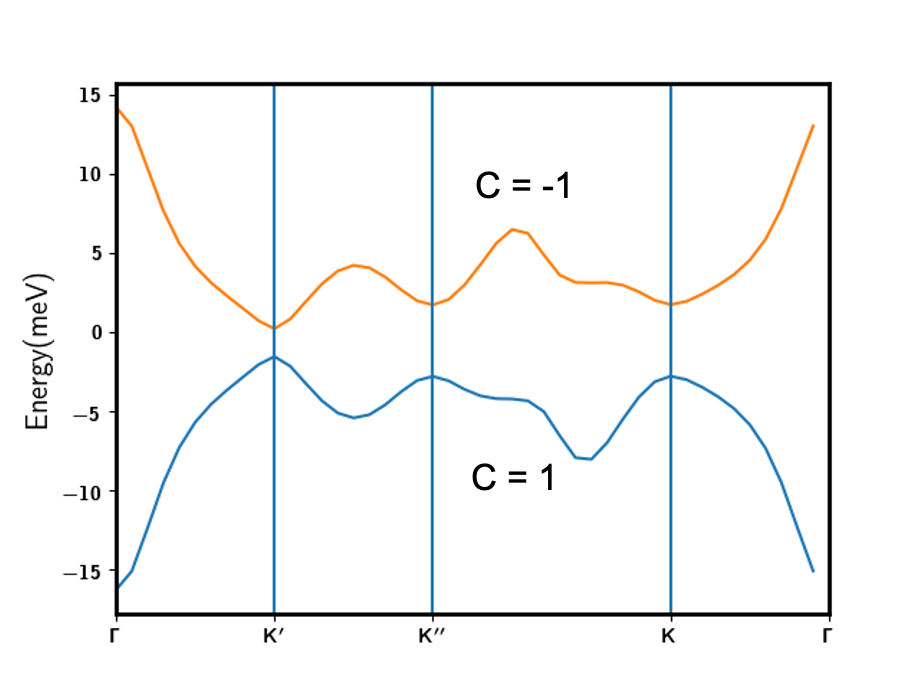}}
 
\caption{Dispersion for (a)Case 1 and (b)Case 2. $\theta_G = 1.2^\circ$.}
\label{fig:dispersion}
\end{figure}

\begin{figure*}[]
\subfloat[]{
  \includegraphics[width = .3\textwidth]{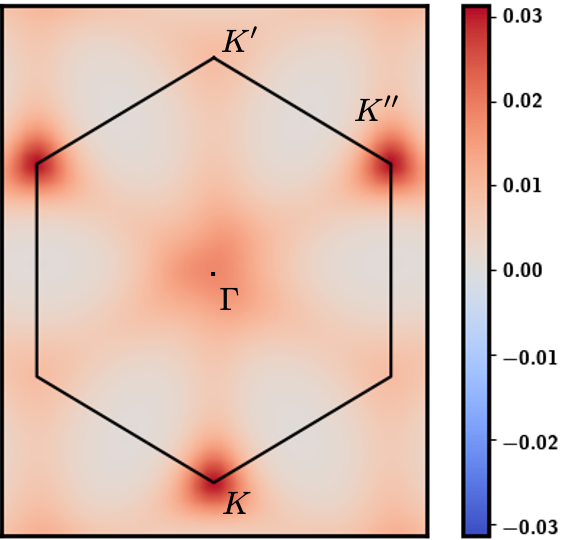}}
\subfloat[]{
  \includegraphics[width = .3\textwidth]{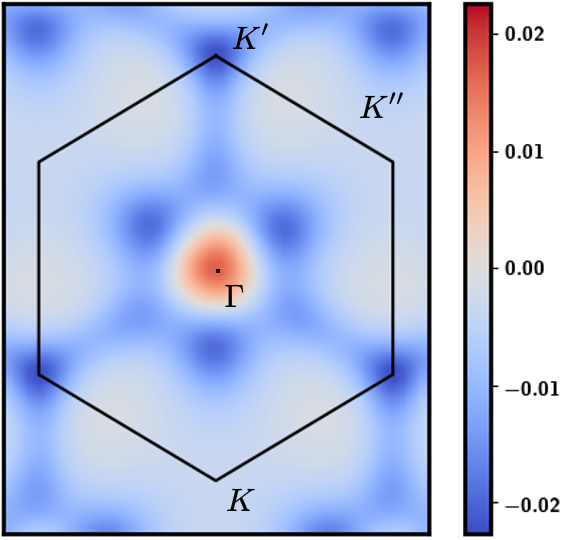}}
\subfloat[]{
  \includegraphics[width = .3\textwidth]{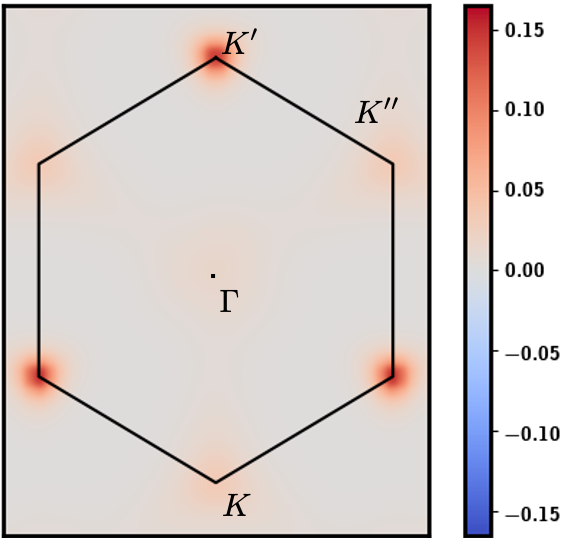}}
 
\caption{Berry curvature distribution of valence band for (a) $m_z$ only (b)Case 1 and (c)Case 2. $\theta_G = 1.2^\circ$. The black line is the boundary of the first BZ and the black dot is the $\Gamma$ point. }
\label{fig:berry}
\end{figure*}

We plot the dispersion relation of the valence and conduction bands near charge neutrality at $\theta_G = 1.2^\circ$ in Fig.\ref{fig:dispersion} along a path in mini BZ for both case 1 and case 2. The contribution to the gap of the momentum dependent terms $V(\vec{Q}_i)$'s depends strongly on how hBN is aligned with TBLG.  Further calculation shows that in case 1, the two bands near charge neutrality has Chern number $\pm 1$, while in case 2, the Chern numbers get reversed. The distribution of Berry curvature of valence band for various cases is plotted in Fig.\ref{fig:berry}. For $\theta_G = 1.15^\circ$, we get the same Chern numbers for case 1 and case 2. The distribution of Berry curvature is also similar to $\theta_G = 1.2^\circ$ (see appendix \ref{appendix}). From the numerical calculation, we demonstrate that the alignment with hBN has a significant effect on the low energy physics of the TBLG system. In particular, the periodic potential induced by hBN cannot be ignored.

\section{Incommensurate Alignment}
\label{sec:incomm}
In general, the alignment between hBN and TBLG is not commensurate. The periodic potential term $V(\vec{Q}_i)$ induces a quasi-periodic potential ralative to the TBLG superlattice. The spectrum of the TBLG will get broadened but since the coupling strength between hBN and graphene is much smaller than the band gaps from the flat bands to the other bands in TBLG system, we can ignore the other bands and only consider an effective description for the flat bands.

The extra hBN layer breaks the $C_2\mathcal{T}$ symmetry of the TBLG system such that there is no obstruction of constructing localized Wannier orbitals using the two bands near charge neutrality in one valley. The low energy effective tight binding model is obtained in two steps. First, take $m_z$ term in $H_V$ together with $H_{TBLG}$ and construct localized Wannier orbitals for the two bands in one valley. Second, project the $V(\vec{Q}_i)$ terms to the active bands and transform to Wannier basis. 

We use projection method to obtain Wannier functions.\cite{marzari_maximally_2012} The relationship between Bloch function and Wannier function can be written as,
\begin{equation}
    \phi_{n,\vec{x}_0}^\dag = \frac{1}{\sqrt{N}} \sum_{\vec{k},m} e^{-i\vec{k}\cdot \vec{x}_0}\psi_{m,\vec{k}}^\dag (U_{\vec{k}})_{mn},
\end{equation}
where $\phi_{n,\vec{x}_0}^\dag$ is the creation operator for Wannier orbital labeled by $n$ at position $\vec{x}_{0}$ and $\psi_{m,\vec{k}}^\dag$ is the creation operator for Bloch state and $m \in \{c,v\}$ labels the conduction band and valence band in one valley. $U_{\vec{k}}$ is a unitary matrix, defined as $U_{\vec{k}} = A_{\vec{k}}(A_{\vec{k}}^\dag A_{\vec{k}})^{-1/2}$, where $(A_{\vec{k}})_{mn} = \langle\mu_{m}(\vec{k})|g_n(\vec{k})\rangle$ is the overlap matrix between the Bloch wave function $|\mu_m(\vec{k})\rangle$ and k-space representation of a localized wave function ansatz $|g_n(\vec{k})\rangle$.  In the numerical calculation below, we take $|g_n(\vec{k})\rangle = e^{-\vec{k}^2/32} e^{-i \vec{k}\cdot \vec{x}_0}|\varphi_n\rangle$ so after inverse Fourier transform, $|g_n(\vec{x})\rangle$ is localized near $\vec{x}_0$. $|\varphi_n\rangle$ is a constant vector in $\vec{k}$ space and it is chosen to maximize the singular values of $A_{\vec{k}}$. \cite{Zhang_bridging}

The projected hopping terms and quasi-periodic potential terms can be written as,
\begin{equation}
    \begin{split}
        t_{mn}(\vec{x}_{ij}) =& \frac{1}{N} \sum_{\vec{k}}e^{i \vec{k} \cdot \vec{x}_{ij}} ( U_{\vec{k}}^\dag \epsilon_{\vec{k}} U_{\vec{k}})_{mn}\\
        V_{mn}(\vec{x}_i, \vec{x}_j) =& \frac{1}{N}\sum_{q=1}^3e^{-i \vec{Q}_q\cdot \vec{x}_j} \sum_{\vec{k}}
    e^{i\vec{k}\cdot \vec{x}_{ij}} ( U_{\vec{k}}^\dag f^{q}(\vec{k}) U_{\vec{k}+\tilde{\vec{Q}}_q})_{mn}
    \label{eq:wannier_proj}
    \end{split}
\end{equation}
where $\vec{x}_{ij} = \vec{x}_i-\vec{x}_j$ is the displacement of the two lattice points, $\epsilon_{\vec{k}} = diag\{\epsilon_c(\vec{k}), \epsilon_v(\vec{k})\}$, $\epsilon_{c,v}(\vec{k})$ being the dispersion of the conduction/valence bands. $\tilde{\vec{Q}}_q$ is defined in the first BZ of the TBLG system and is related to $\vec{Q}_q$ by addition of integer multiples of $\vec{G}_1$ and $\vec{G}_2$. $f^q(\vec{k})$ is the form factor, whose matrix element is,
\begin{equation}
    f_{n_1,n_2}^q(\vec{k}) = \sum_{mn}\psi^*_{n_1t, \vec{k}+ m \vec{G}_1 + n\vec{G}_2} (V_q)_{n_1,n_2} \psi_{n_2t,\vec{k}+\vec{Q}_q+m\vec{G}_1+n\vec{G}_2},
\end{equation}
where $\psi_{n_1t, \vec{k}}$ denotes the wave function in $\vec{k}$-space of the $n_1$ sublattice of the top layer graphene and $V_q$ is the coupling matrix of the quasi-periodic potential term, in the form of Eq.\ref{eq:qp}.

The effective Hamiltonian can therefore be written as,
\begin{equation}
\begin{split}
    H_{tb}=&\sum_{ij}(t_{mn}(\vec{x}_{ij}) c^\dagger_{im}c_{jn}+h.c.) \\
    &+\sum_{ij}(V_{mn} (\vec{x}_i,\vec{x}_j) c^\dagger_{im}c_{jn}+h.c.) ,
\end{split}
\end{equation}
where $c_{i}$ lives on the moir\'e lattice formed by the TBLG and we can write $\vec{r}_i = n_i \vec{a_M}_1+m_i \vec{a_M}_2$, where $\vec{a_M}_1 =  a_M(\frac{1}{2},\frac{\sqrt{3}}{2})$ and $\vec{a_M}_2 = a_M(0,1)$. $a_M = \frac{2a}{\sin(\theta_G/2)}$ being the moir\'e lattice constant. 

\begin{figure}
    \centering
    \includegraphics[width=.4\textwidth]{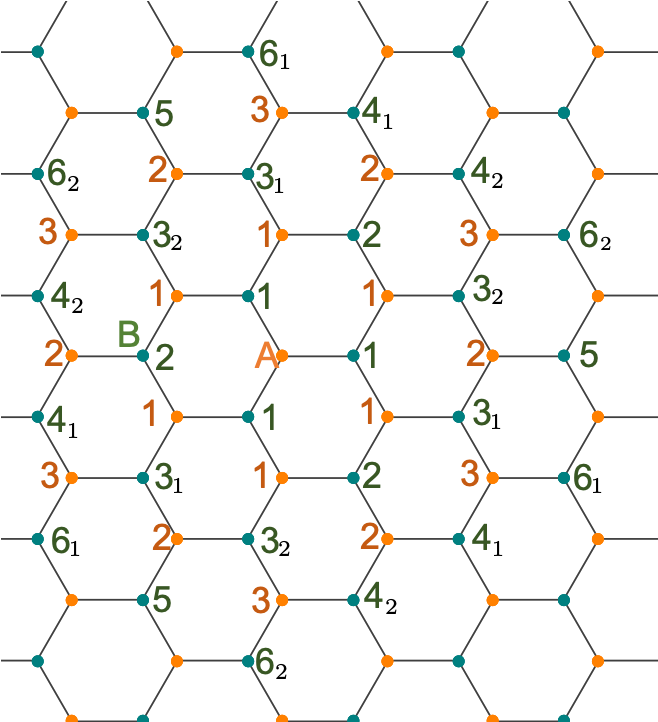}
    \caption{Labels of the hopping terms from site A(orange circle) to other nearby sites. For example, hopping from A to an orange site labelled '2' corresponds to $t_{2AA}$ and from A to an green site labelled $3_1$ corresponds to $t_{3AB1}$.}
    \label{fig:hopping}
\end{figure}

After $U_{\vec{k}}$ is obtained, we get $t_{mn}$'s and $V_{mn}$'s from Eq. \ref{eq:wannier_proj}. Let us consider $t_{mn}$'s first. We find that in order to reproduce the band gap and band structure well, we need to keep the hopping terms up to the third nearest unit cell. (See Table \ref{tab:hopAABB} and Table \ref{tab:hopAB}.) $t_{0AA}$ and $t_{0BB}$ are on-site potentials for site A and B. The meaning of the other labels is explained in Fig.\ref{fig:hopping}. 

\begin{table}[]
    \centering
    \begin{tabular}{|c|c|}
    \hline
    $\theta_G$& $1.2^\circ$\\
    \hline
    $t_{0AA}$&$4.575$ \\
    $t_{0BB}$&$-1.270$ \\
    $t_{1AA}$&$1.547 e^{i(-0.197)\pi}$ \\
    $t_{1BB}$&$-1.613e^{i(-0.188)\pi}$ \\
    $t_{2AA}$&$0.482e^{i(-0.349)\pi}$ \\
    $t_{2BB}$&$-0.452e^{i(0.316)\pi}$ \\
    $t_{3AA}$&$0.506e^{i(-0.134)\pi}$ \\
    $t_{3BB}$&$-0.521e^{i(-0.13)\pi}$ \\
    \hline
    \end{tabular}
    \caption{Hopping between the same type of lattice sites}
    \label{tab:hopAABB}
\end{table}

\begin{table}[]
    \centering
    \begin{tabular}{|c|c|}
    \hline
    $\theta_G$& $1.2^\circ$ \\
    \hline
    $t_{1AB}$&$2.249 e^{i(0.082)\pi}$\\
    $t_{2AB}$&$-1.54 e^{i(0.333)\pi}$\\
    $t_{3AB1}$&$-0.398 e^{i(0.123)\pi}$\\
    $t_{3AB2}$&$0.668 e^{i(-0.304)\pi}$\\
    $t_{4AB1}$&$-0.412 e^{i(0.131)\pi}$\\
    $t_{4AB2}$&$-0.590 e^{i(0.133)\pi}$\\
    $t_{5AB}$&$-0.270  e^{i(0.314)\pi}$\\
    $t_{6AB1}$&$-0.320 e^{i(-0.142)\pi}$\\
    $t_{6AB2}$&$-0.165 e^{i(0.289)\pi}$\\
    \hline
    \end{tabular}
    \caption{Hopping between AB lattice sites}
    \label{tab:hopAB}
\end{table}

\begin{figure}
    \centering
    \includegraphics[width  = 0.4\textwidth]{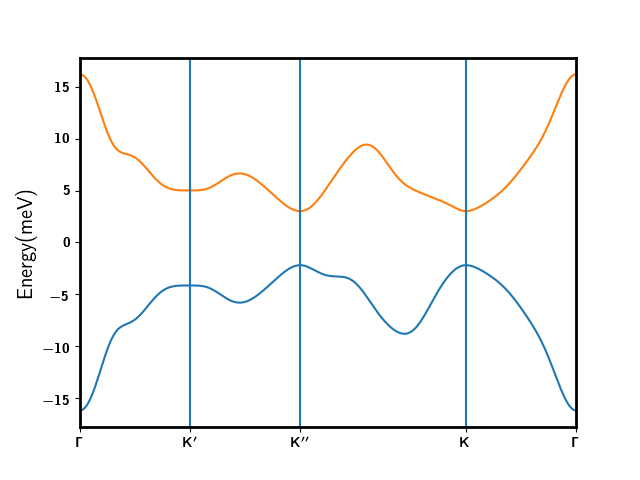}
    \caption{Dispersion of the effective tight binding model}
    \label{fig:TBdispersion}
\end{figure}

Without the quasi-periodic terms, the dispersion of $H_{tb}$ is plotted in Fig.\ref{fig:TBdispersion}. The valence and conduction bands have Chern number $\pm 1$ respectively.

Let us consider the $V_{mn}$ terms. There are two effects of the twist angle between hBN and graphene $\theta_{BN}$ on the tight-binding Hamiltonian: one is the change of the $\vec{Q}$'s and the other is the change of projected amplitude of the quasi-periodic potential terms. We study two different $\theta_{BN}$'s numerically: $\theta_{BN}=0^\circ$ and $0.8^\circ$. We project the quasi-periodic terms to the Wannier orbitals and calculate $V_{mn}(\vec{x}_i, \vec{x}_j)$. We find that although the amplitude of the quasi-periodic terms decays with $|\vec{x}_i - \vec{x}_j|$ but within the 4th nearest neighbor of is on the order of $\sim 1$ meV, which is comparable to the hopping terms. We keep up to the 4th nearest neighbor quasi-periodic terms in the following calculations due to the comparable magnitude of them.

\begin{figure}
\subfloat[\label{fig:DOS0}]{
  \includegraphics[width = .4\textwidth]{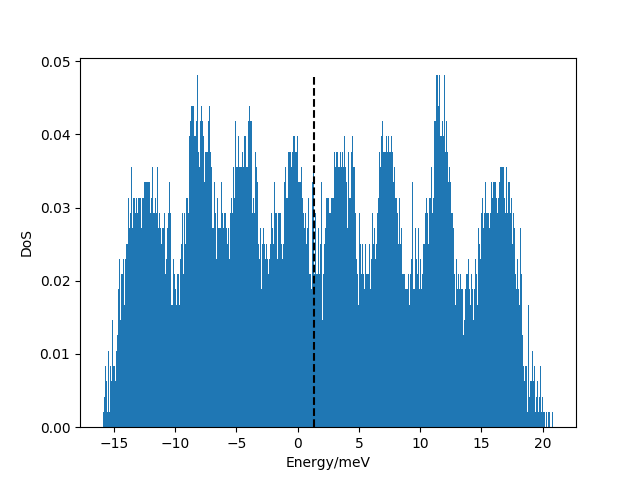}}
  
\subfloat[\label{fig:DOS0.8}]{
  \includegraphics[width = .4\textwidth]{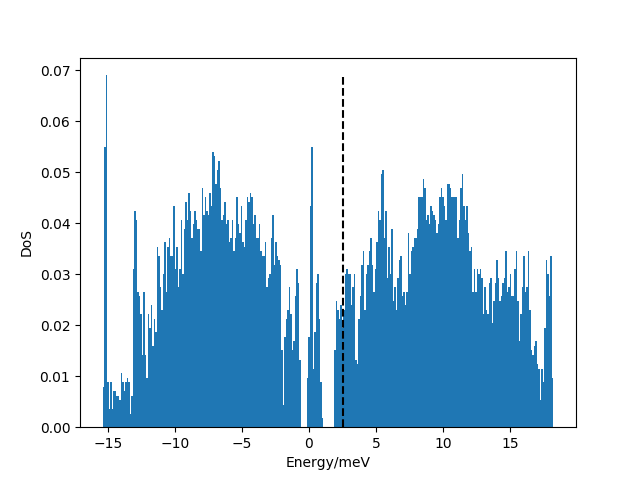}}
  \caption{Density of states at (a) $\theta_{BN} = 0^\circ$ and (b) $\theta_{BN} = 0.8^\circ$ for a $71\times71$ grid in $\vec{k}$-space. The dotted line indicates the energy of the middle state in the spectrum.}
  \label{fig:DOS}
\end{figure}

We plot the density of states for $\theta_{BN} = 0^\circ$ and $\theta_{BN} = 0.8^\circ$ in Fig.\ref{fig:DOS}. At both angles, there are some small peaks but those peaks don't form isolated sub-bands due to the incommensurate nature of the quasi-periodic term. There are 8 main peaks for $\theta_{BN} = 0^\circ$ which can be explained by the commensurate approximation. We can always find a sequence of rational numbers to approximate an irrational number by means of continued fraction expansion. Let us write $\vec{Q}_{1n} = \frac{s_n}{r_n}\vec{G}_1 + \frac{t_n}{r_n}\vec{G}_2$, where $s_n,t_n,r_n$ are integers and $lim_{n\rightarrow\infty} Q_{1n} = \vec{Q}_1$. For each finite $n$, the BZ is folded into a mini-BZ with reciprocal lattice vectors $(\vec{G}_{1n},\vec{G}_{2n}) = ( \vec{G}_1/r_n,\vec{G}_2/r_n)$ with $2r_n^2$ orbitals at each $\vec{k}$ point. For $\theta_{BN} = 0^\circ$, we get $\vec{Q}_1(\theta_{BN} = 0^\circ) \approx 0.49 \vec{G}_1 + 0.47 \vec{G}_2$ so the first order approximation is $(s_1,t_1,r_1) = (1,1,2)$. Thus, there are roughly eight "bands". The main difference between the two $\theta_{BN}$'s is that for $\theta_{BN} = 0^\circ$, the spectrum is gapless near charge neutrality while it is gapped for $\theta_{BN} = 0.8^\circ$ and the gap size is reduced to $\sim 1$meV compared to $\sim 7$meV without quasi-peridoic potential. 

\begin{figure*}
\subfloat[\label{fig:IPR0}]{
  \includegraphics[width = .4\textwidth]{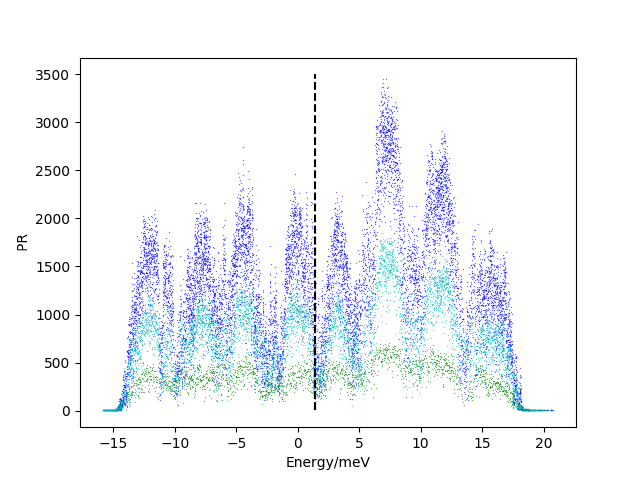}}
\subfloat[\label{fig:IPR0.8}]{
  \includegraphics[width = .4\textwidth]{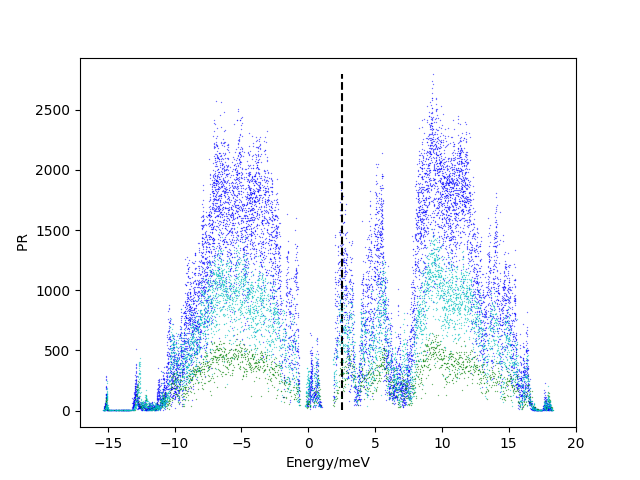}}
  \caption{PR at (a) $\theta_{BN} = 0^\circ$ and (b) $\theta_{BN} = 0.8^\circ$ for different states throughout entire energy spectrum. The dotted line indicates the energy of the middle state in the spectrum. Different color indicates different system size. From bottom to top: $N=30,50,70$ and the system size is $N\times N$ in $\vec{k}$-space.}
  \label{fig:PR}
\end{figure*}
\begin{figure*}
\subfloat[\label{fig:sigma0}]{
  \includegraphics[width = .4\textwidth]{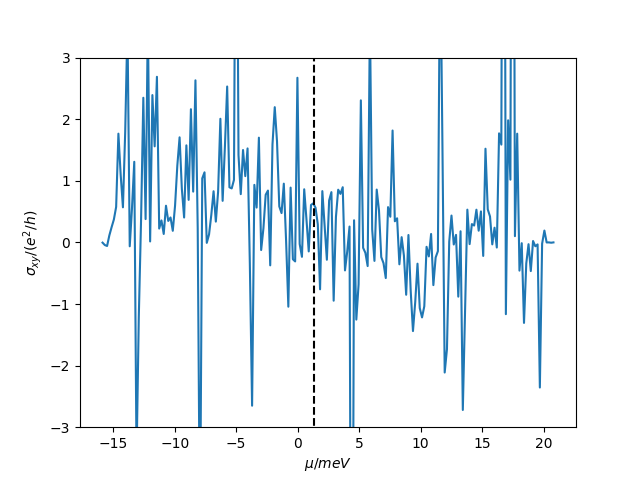}}
\subfloat[\label{fig:sigma0.8}]{
  \includegraphics[width = .4\textwidth]{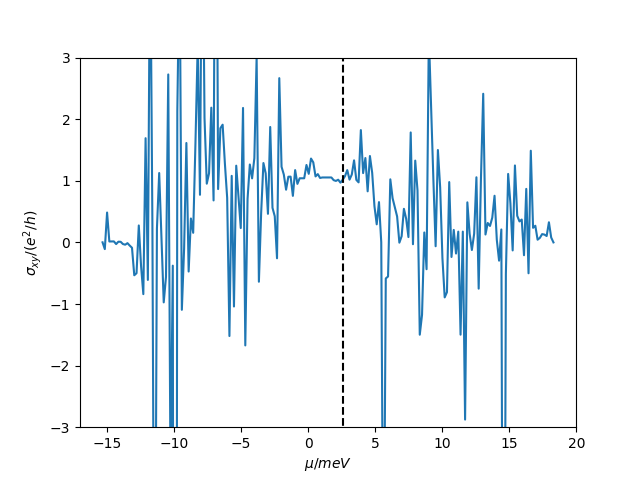}}
  \caption{$\sigma_{xy}$ vs. chemical potential at (a) $\theta_{BN} = 0^\circ$ and (b) $\theta_{BN} = 0.8^\circ$ for a $50\times50$ grid in $\vec{k}$-space. The dotted line indicates the energy of the middle state in the spectrum.}
  \label{fig:sigmaxy}
\end{figure*}

We then study the localization properties of the states. We introduce PR(participation ratio) in $\vec{k}$-space, $PR = \frac{(\sum_{\vec{k}} |\psi_{\vec{k}}|^2)^2}{\sum_{\vec{k}} |\psi_{\vec{k}}|^4}$. States that are localized in real space are extended in momentum space and so we expect $|\psi_{\vec{k}}|\sim 1/N$, where $N\times N$ is the system size in $\vec{k}$-space. Thus PR $\sim N^2$ for localized states and PR $\sim$ constant for extended states. From Fig.\ref{fig:PR}, we find that the PR share similar feature as the density of states, which means that near the dips of density of states(DOS), there are more extended states while near the peaks of DOS, there are more localized states. Localized states in $\vec{k}$-space are extended in real space. Thus, we indeed get metallic behavior near charge neutrality for $\theta_{BN}=0^\circ$ and mobility edges exist. 

The density of states indicates that the alignment of hBN has a strong effect on the low energy physics especially near charge neutrality. We then calculate the Hall conductivity $\sigma_{xy}$ using Kubo formula for one valley and one spin species to further address the difference in electrical transport. In the full many body system, this is the Hall conductivity obtained (within Hartree-Fock) if the system is spontaneously fully spin and valley polarized at the filling considered. Thus at $3/4$ filling of the conduction band (as appropriate for the experiments of Refs. \onlinecite{sharpe_emergent_2019,serlin_intrinsic_2020}, within a Hartree-Fock treatment, full spin-valley polarization leads to full hole filling of one of the Chern bands. This corresponds to placing the effective chemical potential of the Hartree-Fock bands at charge neutrality.   We plot $\sigma_{xy}$ as a function of the effective chemical potential in Fig.\ref{fig:sigmaxy}. For $\theta_{BN} = 0.8^\circ$, $\sigma_{xy}$ is quantized to $\sim 1$ if the chemical potential is slightly below charge neutrality while for $\theta_{BN} = 0^\circ$ it is not quantized.

\section{Chern bands with quasi-periodic potential: a toy model}
\label{sec:toy}
The natural occurrence of topological bands and a quasi-periodic potential in TBLG/hBN discussed in previous sections leads to a number of interesting theoretical questions. For ordinary non-topological bands, the question of how different the effects of a quasiperiodic potential are as compared to a random potential on the electronic wave functions has begun to be addressed in recent years. \cite{devakul_anderson_2017,rossignolo_localization_2019,huang2019moire} Here we are interested instead in similar questions when, in addition, the bands are topological. Within a free fermion theory, what is the behavior of the conductivity as a function of chemical potential? As part of addressing this question, it is important to understand in the first place how to couple in an external vector potential to the electrons in the topological band which is itself a subtle question, as we shall see. 

Here we address these questions within a simple context. 
Let us consider a system with a Chern band and add a quasi-periodic potential to it. If the strength of the quasi-periodic potential is much smaller than the band gaps between the topological band considered and all the other bands, the minimal approach is to project the Hamiltonian to the low energy Chern band. Since there is Wannier obstruction, a tight-binding model in real space is not possible. Thus, we write the effective Hamiltonian in momentum space. For simplicity, assume that we have a flat band to begin with. In momentum space, the Hamiltonian can be written as,
\begin{equation}
    H = \sum_{\vec{k}}\sum_{\vec{Q}_i} c_{\vec{k}}^\dag V(\vec{Q}_i) c_{\vec{k}+\vec{Q}_i} \lambda(\vec{k},\vec{k}+\vec{Q}_i)+h.c.,
\end{equation}
where $\lambda(\vec{k},\vec{k}+\vec{Q}_i)$ is the form factor and $\vec{Q}_i$'s are the reciprocal vectors for the quasi-periodic potential. For each $\vec{k}$, the Hamiltonian can be viewed as a tight-binding model in momentum space with a lattice generated by $\vec{Q}_i$'s. For $\vec{Q}_i$'s that are incommensurate with the original reciprocal lattice vectors $\vec{G}_{1,2}$ that generate the Brillouin zone, we expect $\vec{k}+n_1 \vec{Q}_1 +n_2 \vec{Q}_2$ mod $(m_1 \vec{G}_1+ m_2\vec{G}_2)$ ($n_{1,2}$ and $m_{1,2}$ are integers) to be dense in the first Brillouin zone. In this case, we only need to consider one lattice that is generated by $\vec{k} +n_1 \vec{Q}_1 +n_2 \vec{Q}_2$ with a fixed $\vec{k}$. To keep contact with moir\'e graphene, we will let $\vec{Q}_{1,2}$ generate a triangular lattice but similar discussion can be carried out on any lattice.

For trivial bands, one can take $\lambda(\vec{k},\vec{k}+\vec{Q}_i) = 1$. In this case, the eigenvectors are plane waves in $\vec{k}$ space and therefore they are localized in real space. Thus at large quasi-periodic potential strength for trivial bands, there is always localization.

For a Chern band, on the other hand, the form factor is non-trivial. For small $|\vec{Q}|$, it can be written as $\lambda(\vec{k},\vec{k}+\vec{Q}_i) = F(\vec{k},\vec{k}+\vec{Q}_i)e^{-i \int_{\vec{k}}^{\vec{k}+\vec{Q}_i}\vec{A}(\vec{q}) \cdot d\vec{q}}$, where $F(\vec{k},\vec{k}+\vec{Q}_i)$ is real and positive and the path of the integral is taken to be a straight line from $\vec{k}$ to $\vec{k}+\vec{Q}_i$. To simplify the problem, we assume homogeneous Berry curvature and further let $F(\vec{k},\vec{k}+\vec{Q}_i)=1$ for now. Then the Hamiltonian is equivalent to a tight-binding model in a uniform perpendicular magnetic field. We choose Landau gauge such that $\vec{A}(\vec{k}) = (-B k_y, 0, 0)$. $B$ is proportional to Chern number and the magnetic flux is in general not rational. For this choice of gauge, $\lambda(\vec{k},\vec{k}') = e^{i\frac{B}{2}(k_x'-k_x)(k_y'+k_y)}$. The Hamiltonian can be written as,
\begin{equation}
\begin{split}
        H = &\sum_{\vec{k}} (V_1 c_{k_x,k_y}^\dag c_{k_x+1,k_y} e^{i B k_y} \\
        &+V_2 c_{k_x,k_y}^\dag c_{k_x-1/2,k_y+\sqrt{3}/2} e^{-i\frac{B}{2}(k_y + \frac{\sqrt{3}}{4})} \\
        &+ V_3 c_{k_x,k_y}^\dag c_{k_x-1/2,k_y-\sqrt{3}/2}e^{-i\frac{B}{2}(k_y - \frac{\sqrt{3}}{4})}+h.c.),
        \label{eq:toymodel_1band}
\end{split}
\end{equation}
where the lattice spacing is set to be $1$ and $V_{1,2,3}$ are taken to be real. The Hamiltonian in Eq.\ref{eq:toymodel_1band} is a special case of that considered in \cite{han_critical_1994}. Following the arguments in \cite{han_critical_1994}, we can perform a Fourier transform along $k_x$ direction. The 2D model is then equivalent to a 1D lattice model with quasi-periodic(QP) potential. One can write $k_y = k_y^0 + n\frac{\sqrt{3}}{2}$, where $n$ is an integer. Note that we assume $\vec{k}$-lattice is dense in the original BZ. We can take $k_y^0$ to be $0$. The flux quanta is $\Phi = \frac{\sqrt{3}B}{4\pi}$.  Thus the 1D lattice model with QP potential is,
\begin{equation} 
    \begin{split}
        E\phi_n = 2V_1 \cos(2\pi n \Phi + \nu) \phi_n + A_{n,n+1} \phi_{n+1} + A_{n,n-1} \phi_{n-1},
    \end{split}
    \label{eq:1d}
\end{equation}
where $A_{n,n+1} = V_2 e^{-i[\pi\Phi(n+\frac{1}{2})]-i\frac{\nu}{2}}+ V_3e^{i[\pi\Phi(n+\frac{1}{2})]+i\frac{\nu}{2}}$ and $A_{n,n-1} = A_{n-1,n}^*$. $\psi(k_x,k_y)$ is the eigenfunction for the Hamiltonian \ref{eq:toymodel_1band} at energy $E$, $\psi(k_x,k_y) = e^{i k_x \nu} \phi(k_y)$ and $\phi_n\equiv \phi(k_y^0+n\frac{\sqrt{3}}{2})$.

Depending on the relative strength of $V_{1,2,3}$, $\phi(k_y)$ can be either localized or extended in $k_y$ space. If there is $C_3$ rotational symmetry, which corresponds to $V_1=V_2=V_3$, the 1D system is at the critical point of the localization transition and thus the eigenstates are not localized in real space, which is different from the trivial band case.

If the $C_3$ symmetry is broken, we get the Lyapunov exponent (inverse of the localization length) by considering three different gauge choices, i.e., along the three axes of the triangular lattice\cite{Thouless_bandwidth,han_critical_1994} and the Lyapunov exponent $\lambda(E;V_1,V_2,V_3)$ for $\phi_n$ in Eq.\ref{eq:1d} is,
\begin{equation}
    \lambda(E;V_1,V_2,V_3) = ln\left(\frac{|V_1|}{|V_3|}\right),
\end{equation}
if $|V_1|\geq|V_3|\geq|V_2|$. And
\begin{equation}
    \lambda(E;V_1,V_2,V_3) = ln\left(\frac{|V_1|}{|V_2|}\right),
\end{equation}
if $|V_1|\geq|V_2|\geq|V_3|$. $ \lambda(E;V_1,V_2,V_3) = 0$ otherwise.

Even though we get localized or extended $\psi(k_x,k_y)$, depending on the choices of $V_{1,2,3}$, we still need to address the question of what effect it will have on the physical observables. Thus, we study the DC transport of the system in the following.

\begin{figure}
    \centering
    \includegraphics[width=0.4\textwidth]{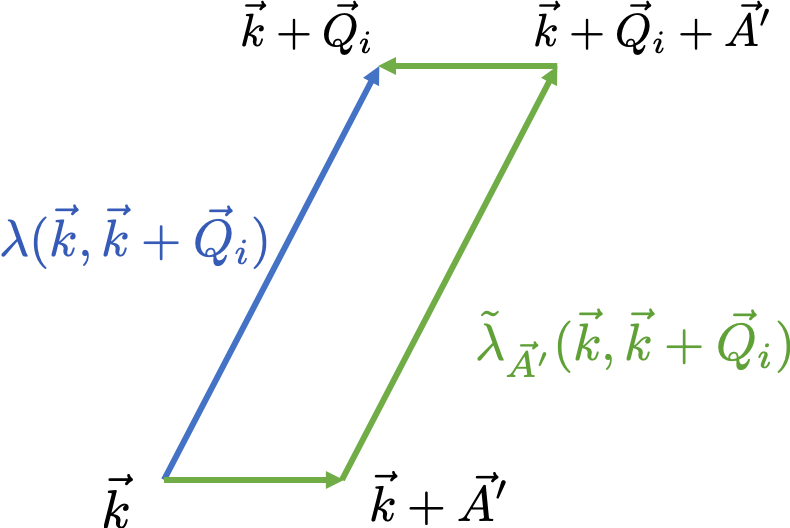}
    \caption{Schematic of the line integrals in Eq.\ref{eq:peierls}}
    \label{fig:contour}
\end{figure}

In the trivial case, all the states are localized so we expect the conductance to vanish. In the topological case, we need to couple the tight-binding Hamiltonian in k-space to external electric field. First we need to obtain current operators in the presence of external electric field. The strategy is to apply a probe vector potential $\vec{A'}$ and $J_\mu = -\frac{\partial H[\vec{E},\vec{A'}]}{\partial A'_\mu}|_{A'_\mu\rightarrow 0}$. The vector potential $\vec{A'}$ will "shift" the momenta $\vec{k}$. We have to be careful about what we mean by "shift". In comparison to the trivial band, there is a gauge structure in $k$-space. If we change $c_{\vec{k}}$ to $c_{\vec{k}}e^{i\theta(\vec{k})}$ and $\vec{A}(\vec{k})$ to $\vec{A}(\vec{k})+\partial_{\vec{k}} \theta(\vec{k})$, where $\theta(\vec{k})$ is a differentiable function in $\vec{k}$, $c_{\vec{k}}^\dag c_{\vec{k}+\vec{Q}} \lambda(\vec{k},\vec{k}+\vec{Q})$ is invariant. In order to keep the gauge invariance of the theory, we cannot simply replace $\lambda(\vec{k},\vec{k}')$ by $\lambda(\vec{k}+\vec{A}',\vec{k}'+\vec{A}')$. The only gauge-invariant deformation of the form factor $\lambda(\vec{k},\vec{k}')$ is to attach a small plaquette with Berry curvature as flux. This generalizes the idea of Peierls substitution. Again, let us only consider the phase factor in the form factor for now. The 
gauge invariant change in the form factor $\lambda(\vec{k},\vec{k}+\vec{Q}_i)$ by shifting the momentum by  $\vec{A}'$ is,
\begin{equation}
   \tilde{\lambda}_{\vec{A}'}(\vec{k},\vec{k}+\vec{Q}_i)-\lambda(\vec{k},\vec{k}+\vec{Q}_i) = \lambda(\vec{k},\vec{k}+\vec{Q}_i) e^{i B (\vec{Q}_i \times \vec{A}')\cdot \hat{e}_z },
   \label{eq:peierls}
\end{equation}
where $\tilde{\lambda}_{\vec{A}'}(\vec{k},\vec{k}+\vec{Q}_i)$ denotes the shifted form factor and it is defined as the Wilson loop of the Berry connection along the green curve in Fig.\ref{fig:contour}. $\hat{e}_z$ is the directional vector along $z$-axis.

The way that we construct the coupling to the external vector potential in Eq.~\ref{eq:peierls} seems to rely on the specific form of the form factor $\lambda(\vec{k},\vec{k}+\vec{Q})$. For a generic form factor, we also construct the gauge coupling (see Appendix \ref{appendixB} for details), which agrees with Eq.~\ref{eq:peierls}.

The tight-binding Hamiltonian coupled to the probe vector potential and external electric field can be written as,
\begin{equation}
    \begin{split}
        &H[\vec{E},\vec{A}']  = \sum_{\vec{k}} (V_1 c_{k_x,k_y}^\dag c_{k_x+1,k_y} e^{i B k_y}e^{iB A'_y} \\
        &+V_2 c_{k_x,k_y}^\dag c_{k_x-1/2,k_y+\sqrt{3}/2} e^{-i\frac{B}{2}(k_y + \frac{\sqrt{3}}{4}+A'_y)} e^{-i B \frac{\sqrt{3}}{2} A'_x} \\
        &+ V_3 c_{k_x,k_y}^\dag c_{k_x-1/2,k_y-\sqrt{3}/2}e^{-i\frac{B}{2}(k_y - \frac{\sqrt{3}}{4}+A'_y)}e^{i B \frac{\sqrt{3}}{2} A'_x}\\
        & +h.c.)\\
        &+ V(\vec{E},\vec{A}')\\
    \end{split}
\end{equation}
$V(\vec{E},\vec{A}') = \int d^2 x \vec{E}\cdot \vec{x} \rho(\vec{x};\vec{A}')$ is the electric potential, where $\rho(\vec{x};\vec{A}')$ is the density operator. In $k$-space, we define the Fourier transformation of $\rho(\vec{x};\vec{A}')$ as $\rho(\vec{q};\vec{A}')$ and
\begin{equation}
\begin{split}
        \rho(\vec{q};\vec{A}') &= \sum_{\vec{k}}c_{\vec{k}}^\dag c_{\vec{k}+\vec{q}}\lambda(\vec{k},\vec{k}+\vec{q})e^{iB(\vec{q} \times \vec{A}')\cdot \hat{e}_z}\\
        &=\rho(\vec{q};\vec{A}'=0) e^{iB(\vec{q} \times \vec{A}')\cdot \hat{e}_z}.
\end{split}
\end{equation}
Thus the electric potential can be written as,
\begin{equation}
\begin{split}
         &V(\vec{E},\vec{A}') = \iint d^2 q d^2 x \vec{E}\cdot \vec{x} \rho(\vec{q};\vec{A}'=0) e^{-i \vec{q}\cdot (\vec{x}- B \vec{A}'\times\hat{e}_z)}\\
         &=V(\vec{E},\vec{A}'=0)
        +B\vec{E}\cdot (\vec{A}'\times\hat{e}_z) \rho(\vec{q}=0;\vec{A}'=0)
\end{split}
\end{equation}
to linear order in $B$, 
and the probe vector field $\vec{A}'$ only couples to the second term above.

We can do a sanity check of the above expression. Suppose there's no quasi-periodic potential, we can get $J_{x,0} = -1/S\sum_{\vec{k}} c_{\vec{k}}^\dag c_{\vec{k}}B E_y$ and $J_{y,0} = 1/S\sum_{\vec{k}} c_{\vec{k}}^\dag c_{\vec{k}}B E_x$, where $S$ is the total area of the system. For a fully-filled Chern band with Chern number $C$, $B = \frac{2\pi C}{A_{BZ}} = \frac{C A_{cell}}{2\pi}$, where $A_{BZ}$($A_{cell}$) is the area of the Brillouin zone(unit cell). Thus the Hall conductivity $\sigma_{xy} = \frac{B}{A_{cell}} = \frac{C}{2\pi}$ is quantized. 

Now that we have some confidence in the Peierls substitution, let us take the quasi-periodic potential into account. The current density operators can be expressed as,
\begin{equation}
    \begin{split}
         J_{x} = &\frac{1}{S}\sum_{\vec{k}} (i\frac{\sqrt{3}}{2} V_2 B  c_{k_x,k_y}^\dag c_{k_x-1/2,k_y+\sqrt{3}/2} e^{-i\frac{B}{2}(k_y + \frac{\sqrt{3}}{4})}  \\
        &- i\frac{\sqrt{3}}{2}V_3 B c_{k_x,k_y}^\dag c_{k_x-1/2,k_y-\sqrt{3}/2}e^{-i\frac{B}{2}(k_y - \frac{\sqrt{3}}{4})}\\
        & +h.c.)-\frac{1}{S}\sum_{\vec{k}} c_{\vec{k}}^\dag c_{\vec{k}}B E_y\\
        J_{y}= -& \frac{1}{S}\sum_{\vec{k}} (i B V_1 c_{k_x,k_y}^\dag c_{k_x+1,k_y} e^{i B k_y} \\
        &-i\frac{B}{2}V_2 c_{k_x,k_y}^\dag c_{k_x-1/2,k_y+\sqrt{3}/2} e^{-i\frac{B}{2}(k_y + \frac{\sqrt{3}}{4})}\\
        &-i\frac{B}{2} V_3 c_{k_x,k_y}^\dag c_{k_x-1/2,k_y-\sqrt{3}/2}e^{-i\frac{B}{2}(k_y - \frac{\sqrt{3}}{4})}\\
        & +h.c.))+\frac{1}{S}\sum_{\vec{k}} c_{\vec{k}}^\dag c_{\vec{k}}B E_x
    \end{split}
    \label{eq:current}
\end{equation}
Let us consider $E_x = E$ and $E_y=0$. The transport properties are the same as a tight-binding model in real space if we view $k_x$ as $y$-coordinate, $k_y$ as $x$-coordinate. By making this mapping, we transform the problem of non-trivial Berry curvature in $\vec{k}$-space to the problem of a real-space tight binding model under perpendicular magnetic field. In the original model, $\vec{E}$ is along $x$-direction while $\vec{E}$ is along $y$-direction in the real-space model. This is the same as what happens in lowest Landau level(LLL). Indeed, if we
view the wave functions for the LLL as wave functions for the flat Chern band at $C=1$ and calculate the form factors, by comparing with \cite{GMP}, we get the drift current exactly in the form of Eq.\ref{eq:current}.

For commensurate flux $\Phi=2p/q$, translational symmetry is restored and the energy spectrum is divided into $q$ magnetic sub-bands. For simplicity, we only consider zero temperature. $\sigma_{xx} =0$ if the chemical potential is within band gaps. $\sigma_{xy}$ can be obtained through TKNN formula\cite{TKNN},
\begin{equation}
    \sigma_{xy} = \sum_{m}\int\frac{d^2\nu}{(2\pi)^2} f_{m}(\vec{\nu})\left[\mathcal{F}_{m}(\vec{\nu}) + \frac{C}{2\pi}\right],
    \label{eq:sigma_xy}
\end{equation}
where $m\in\{1,\ldots,q\}$ is the index of magnetic bands, $f_m(\vec{\nu})$ is the Fermi-Dirac distribution. $\vec{\nu}$ takes value with the magnetic Brillouin zone $\nu_x \in (-\pi,\pi]$ and $\nu_y \in(-\pi/q,\pi/q]$. $\mathcal{F}_{m}(\vec{\nu})+ \frac{C}{2\pi}$ is the total Berry curvature of the states at $\vec{\nu}$ of the $m$-th band, where the first term takes care of the contribution of the QP potential and the second term comes from the background Berry curvature. 

Note that $\sigma_{xy}$ is quantized although it is not obvious from the expression in Eq.\ref{eq:sigma_xy}. Following Ref.\cite{TKNN,Dana}, Eq. \ref{eq:sigma_xy} is reduced to $\sigma_{xy} = \frac{m C}{2\pi}$, where $m$ is an integer that satisfies the Diophantine equation $-p\Delta n +  q m = 1$, where $\Delta n$ is an integers.

For incommensurate flux, by mapping to the real-space model, $\sigma_{xx} = 0$ if the filled states are localized along $k_y$-direction, which is the case for $|V_1|> max(|V_2|,|V_3|)$. Otherwise, if $|V_1|< max(|V_2|,|V_3|)$, all states are extended.

Let us consider $\sigma_{xy}$ next. If the chemical potential is within the gap and the gap is not closed if we continuously tune the value of $\Phi$ from an irrational number to a nearby rational number $\Phi_0 = 2p/q$, the value of $\sigma_{xy}$ is then completely determined by the Berry curvature of the filled bands at $\Phi_0$. The exact value of $\Phi_0$ is determined by the details of the energetics.

So far, we have only considered flat bands. We can further include the kinetic terms $\epsilon_{\vec{k}}c_{\vec{k}}^\dag c_{\vec{k}}$. $\epsilon_{\vec{k}}$ is the dispersion, which is a periodic function in $\vec{k}$ and $\epsilon_{\vec{k}} = \epsilon_{\vec{k}+m\vec{G}_1+n\vec{G}_2}$, where $\vec{G}_{1,2}$ are the reciprocal lattice vectors and $m,n$ are integers. Note that $\vec{G}_{1,2}$ are in general not commensurate with the reciprocal vectors of the quasi-periodic potential $\vec{Q}_{1,2}$ so the kinetic terms act as on-site ``quasi-periodic'' terms of the tight-binding Hamiltonian in $\vec{k}$-space. Roughly speaking, whether an eigenstate is localized or extended is given by the competition between hopping terms and on-site quasi-periodic(QP) potential terms, i.e., the competition between the energy scales of the band width and of the on-site QP potential.  We have shown that Berry curvature plays a role of magnetic field in $\vec{k}$-space and for a tight-binding model under magnetic field, the energy spectrum can in general develop several sub-bands even if $\Phi$ is irrational, as in Hofstadter's butterfly\cite{hofstadter}. The relevant energy scale for the kinetic energy in $\vec{k}$-space is thus the band-widths of the magnetic sub-bands, which is reduced from the band-width of the same tight-binding model but with no Berry curvature. In this sense, it is ``easier'' to get localized states in $\vec{k}$-space, that is, extended states in real space in a topological band than in a trivial band under on-site QP potential. 

With dispersion and(or) non-uniformity of the Berry curvature taken into account, the Hamiltonian written in $\vec{k}$-space cannot be reduced to an equivalent 1D Hamiltonian and one cannot use Thouless formula to obtain the localization length. Nonetheless, we expect that there is at least one state in the spectrum that is extended or critical in real space due to the non-trivial topology of the original Chern band. We also calculate the inverse participate ratio $\left(IPR = \frac{\sum_{\vec{k}} |\psi_{\vec{k}}|^4 }{(\sum_{\vec{k}} |\psi_{\vec{k}}|^2)^2}\right)$ numerically for different system sizes and find that there are more \emph{non-localized} states (extended or critical) for the non-trivial Berry curvature case than for the vanishing Berry curvature case (see Appendix\ref{sec:IPR} for details).

We can take one-step further towards the TBLG aligned with hBN system by considering two flat topological bands with Chern number $\pm 1$. For illustration purpose, we only consider a square lattice; the quasi-periodic potential only contains the lowest harmonics and the system has $C_4$ rotational symmetry. The quasi-periodic potential also only acts within the same band and there is a inter-band mixing term. The Hamiltonian thus can be written as,
\begin{equation}
\begin{split}
        H_{\pm} = &\frac{\Delta}{2} \sum_{\vec{k}} \left(c_{\vec{k};+}^\dag c_{\vec{k};+} - c_{\vec{k};-}^\dag c_{\vec{k};-}\right)\\
        &+ V_0 \sum_{\vec{k},\vec{Q}} c_{\vec{k};\pm}^\dag c_{\vec{k}+\vec{Q};\pm} \lambda_{\pm\pm}(\vec{k},\vec{k}+\vec{Q})\\
        &+ V_1 \sum_{\vec{k}} c^\dag_{\vec{k};+}c_{\vec{k};-}\lambda_{+ - }^0(\vec{k})+h.c.,
\end{split}
\label{eq:pmChern_Ham}
\end{equation}
where the subscripts $\pm$ label the different bands. The $V_0$ terms are projected quasi-periodic potential terms and the $V_1$ terms are inter-band hopping between the two $\pm$ Chern bands.
 The form factor $\lambda_{\pm\pm}(\vec{k},\vec{k}+\vec{Q}) = \langle \psi_{\vec{k};\pm}|e^{-i\vec{Q}\cdot \vec{x}} |\psi_{\vec{k}+\vec{Q};\pm}\rangle$, where $|\psi_{\vec{k};\pm}\rangle$ are Bloch states and similarly,  $\lambda_{+-}^0(\vec{k}) = \langle \psi_{\vec{k};+}| \psi_{\vec{k};-}\rangle$. $\Delta$ is set to be positive. For the purpose of illustration, we take $|\psi_{\vec{k};a}\rangle$ to be the same as in the LLL and choose a Landau gauge $\vec{A}_{\pm} = (\mp B y,0)$ such that $\langle \vec{x} |\psi_{\vec{k};\pm}\rangle = \sum_{m} e^{i (\mp mk_y + k_x x+ mB x\mp k_x k_y/B)} \Psi_0(y\pm\frac{k_x+mB}{B}) $(see Appendix \ref{wavefn} for details), where $\Psi_0(y)=(\frac{B}{\pi})^{\frac{1}{4}}e^{-\frac{B y^2}{2}}$. We further let $\vec{Q}_1 = Q(2\pi,0)$ and $\vec{Q}_2 = Q(0,2\pi)$, where $Q$ is an irrational number (Note that in LLL, we can always define the magnetic Brillouin zone so $\vec{Q}_{1,2}$ are aligned with the mBZ reciprocal lattice vectors $\vec{G}_{1,2}$, which are set to be $(2\pi,0)$ and $(0,2\pi)$ here). As elaborated in Appendix \ref{wavefn}, the form factors are,
 \begin{equation}
     \begin{split}
         \lambda_{\pm\pm} (\vec{k},\vec{k}+\vec{Q}) &= e^{\mp i \frac{2k_y Q_x+Q_xQ_y}{2B}-\frac{4\pi^2 Q^2}{4B}}\\
         \lambda_{+-}^0 (\vec{k}) &=\sum_{m=-\infty}^\infty e^{ \frac{2 i k_y}{B}(k_x+mB)-\frac{(k_x+mB)^2}{B}},
     \end{split}
     \label{eq:lambdas}
 \end{equation}
Plugging in the definition of $\vec{Q}_1$ and $\vec{Q}_2$, we have, $\lambda_{\pm\pm}(\vec{k},\vec{k}+\vec{Q}_1) = e^{-\frac{\pi Q^2}{2}\mp i Q k_y}$ and $\lambda_{\pm\pm}(\vec{k},\vec{k}+\vec{Q}_2) = e^{-\frac{\pi Q^2}{2}}$.

One can check that the $c_{\vec{k};\pm}$ bands are topological bands with Chern number $\pm 1$ in two ways. First, by taking derivatives of $\vec{Q}$ in Eq.\ref{eq:lambdas} around $\vec{Q}=0$, one gets uniform Berry curvature of $\pm \frac{1}{B}$ for $c_{\vec{k};\pm}$ bands respectively. Second, since the phases of the Bloch wave functions $|\psi_{\vec{k};\pm}\rangle$ are well-defined in the whole BZ, the integration of Berry curvature over the BZ is reduced to a contour integral of the Berry connection along the boundary of the BZ. We have $|\psi_{(k_x+2\pi,k_y);\pm}\rangle = |\psi_{(k_x,k_y);\pm}\rangle$ and $|\psi_{(k_x,k_y+2\pi);\pm}\rangle = e^{\pm i k_x}|\psi_{(k_x,k_y);\pm}\rangle$ such that $\vec{\mathcal{A}}_{(\pi, k_y);\pm} = \vec{\mathcal{A}}_{(-\pi, k_y);\pm}$ and $\vec{\mathcal{A}}_{(k_x, \pi);\pm} = \vec{\mathcal{A}}_{(k_x, -\pi);\pm} \mp \hat{\mathbf{e}}_{x}$, where $\vec{\mathcal{A}}_{(k_x,k_y);\pm}$ is the Berry connection and $\hat{\mathbf{e}}_{x}$ is the unit vector along $k_x$. Thus the contour integrals of $\vec{\mathcal{A}}_{\vec{k};\pm}$ along the boundary of the BZ yield Chern number $\pm 1$.

If $V_0=0$, the Hamiltonian in Eq.\ref{eq:pmChern_Ham} is block-diagonal in $\vec{k}$ space. By solving the $2\times2$ block, we have the two eigenvalues $\pm \epsilon_{\vec{k}} = \pm \sqrt{\frac{\Delta^2}{4} +V_1^2 |\lambda_{+-}^0(\vec{k})|^2}$ so the system is always gapped if $\Delta\neq 0$ and has a gap that is $\geq \Delta$. The eigenvectors are,
\begin{equation}
    \begin{cases}
    d_{\vec{p};+} &= \sqrt{\frac{\epsilon_{\vec{p}} +\Delta/2}{2\epsilon_{\vec{p}}}}c_{\vec{p};+}+e^{i \theta(\vec{p})} \sqrt{\frac{\epsilon_{\vec{p}} -\Delta/2}{2\epsilon_{\vec{p}}}} c_{\vec{p};-} \\
    d_{\vec{p};-} &= -e^{-i \theta(\vec{p})}\sqrt{\frac{\epsilon_{\vec{p}} -\Delta/2}{2\epsilon_{\vec{p}}}}c_{\vec{p};+}+ \sqrt{\frac{\epsilon_{\vec{p}} +\Delta/2}{2\epsilon_{\vec{p}}}}c_{\vec{p};-},
    \end{cases}
    \label{eq:d_fields}
\end{equation}
where $d_\pm$ are the annihilation operators for eigenstates in $\pm$ energy bands and $ \theta(\vec{p}) =Arg[ \lambda_{+-}^0(\vec{p})]$. We choose the phase factors such that in the limit of $V_1\rightarrow0$, $d_{\vec{p};\pm}\rightarrow c_{\vec{p};\pm}$. Since the $V_1$ term does not close the gap, we expect the $d_{\vec{k};\pm}$ bands to have the same Chern number as the $c_{\vec{k};\pm}$ bands. Note that $\lambda_{+-}^0(\vec{k}) = 0$ at $(k_x,k_y) = (\pm \pi, \pm \frac{\pi}{2})$ and $\theta(\vec{k})$ is not well-defined at these singular points. However, the factors associated with $\theta(\vec{k})$ in Eq.\ref{eq:d_fields} vanish at $(k_x,k_y) = (\pm \pi, \pm \frac{\pi}{2})$ so the $d_{\vec{k};\pm}$ fields can be continuously defined in the whole BZ.

If $V_1=0$, $c_{\vec{k},+}$ and $c_{\vec{k},-}$ bands are decoupled and each one of the bands is a flat band with QP. As we discussed before, the spectrum of each band has fractal structure and the width of the spectrum is of the order of $V_0$. Moreover, from Eq. \ref{eq:pmChern_Ham}, after a partial Fourier transformation along $k_x$, we find that the $V_0$ terms are the same for $c_{\vec{k};+}$'s and $c_{\vec{k};-}$'s. Thus the energy spectra of $\pm$ bands are identical and the "+" bands are shifted with an energy $\Delta$ from the "-" bands, with the same corresponding energy eigenstates. If $V_0 \ll \Delta$, there is a band gap between the two ``fractal'' bands that consist of $c_{\vec{k},\pm}$ degrees of freedom respectively, and we get a Chern insulator at half filling. If $V_0\sim \Delta$, the gap at half filling will close. If $V_0 \gg \Delta$, the fractal bands contributing positive Hall conductivity and negative Hall conductivity almost overlap, resulting in nearly zero Hall conductivity.

Now we take both $V_0$ and $V_1$ into account. If we fix $V_1$ and $\Delta$ and increase $V_0$, the band gap decreases and eventually vanishes. We further calculate the IPR (see Appendix \ref{sec:IPR} for details). We find that there are extended states (in real space) near band edges when the band gap is not closed. Upon increasing $V_0$, after the band gap closes, there are localized states near zero energy. Similar ``leviation'' and ``pair annihilation'' behavior of the extend states is also observed in disordered topological insulators \cite{onoda2007localization,prodan2010entanglement}. If the strength of QP potential further increases, the states near zero energy get de-localized since in the $V_0\gg V_1$ limit, the model reduces to two decoupled AA models at critical points.

\section{Conclusion}
In this paper we showed that when magic-angle twisted bilayer graphene is nearly aligned with  h-BN, the single particle physics is sensitive to the quasiperiodic potential produced by the interference between two moire potentials: one produced by the relative twist of the two graphene layers, and the other produced by the h-BN substrate. The periodic modulation induced by h-BN cannot be treated as a small perturbation due to the narrow bandwidth of the valence and conduction bands. By exact diagonalization, we find that for TBLG twist angle $1.2^\circ$, for alignment angle $\theta_{BN} = 0^\circ$ and $\theta_{BN} =0.8^\circ$, localized states and extend states are both present and there is no clear mobility edge. For $\theta_{BN} = 0^\circ$, the charge gap near neutrality is closed. In the presence of valley polarization (due to interactions), the Hall conductivity $\sigma_{xy}$ is not quantized when $\theta_{BN} = 0^\circ$ while for $\theta_{BN} = 0.8^\circ$, the charge gap is reduced and $\sigma_{xy}$ is quantized. 

In order to study the electron properties of topological bands in the presence of quasi-periodic potential, it is more straightforward to begin with a model in momentum space since the non-triviality is manifest in the form factor. In the limit of flat band and uniform Berry curvature, we find that quasi-periodic potential induces hopping between different momentum, which can be mapped to a tight-binding model coupled to  magnetic field. We discussed localization properties and transort  in such toy models. The next step will be to introduce dispersion and electron-electron interaction, which we leave for future studies.

\section{Acknowledgement}
We thank Anushya Chandran, Zhihuan Dong, David Goldhaber-Gordon, Pablo Jarillo-Herrero, and Ya-Hui Zhang for useful discussions. 
This work was supported by NSF grant DMR-1911666,
and partially through a Simons Investigator Award from
the Simons Foundation to Senthil Todadri. This work was also partly supported by the Simons Collaboration on Ultra-Quantum Matter, which is a grant from the Simons Foundation (651440, TS).

\bibliographystyle{apsrev4-1}
\bibliography{hBN_moire}

\onecolumngrid
\appendix

\section{Numerical results for $\theta_{G} = 1.15^\circ$}
\label{appendix}
\subsection{Perfect Alignment}
\begin{figure}[ht]
\subfloat[]{
  \includegraphics[width = .32\textwidth]{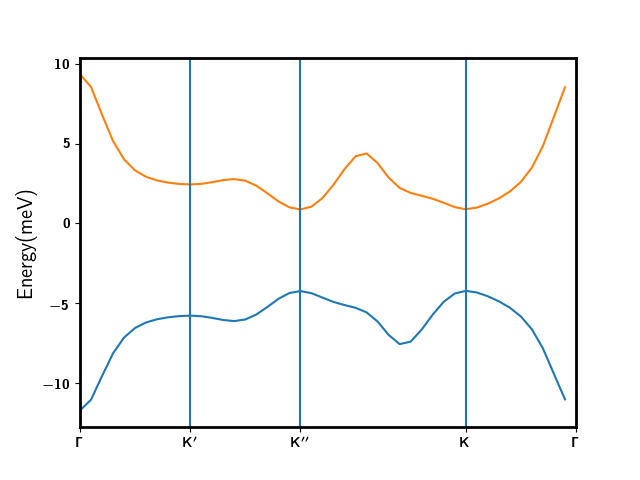}}
\subfloat[]{
  \includegraphics[width = .32\textwidth]{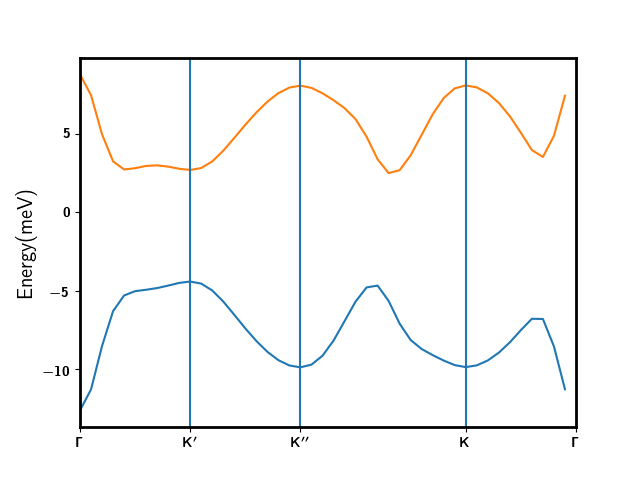}}
\subfloat[]{
  \includegraphics[width = .32\textwidth]{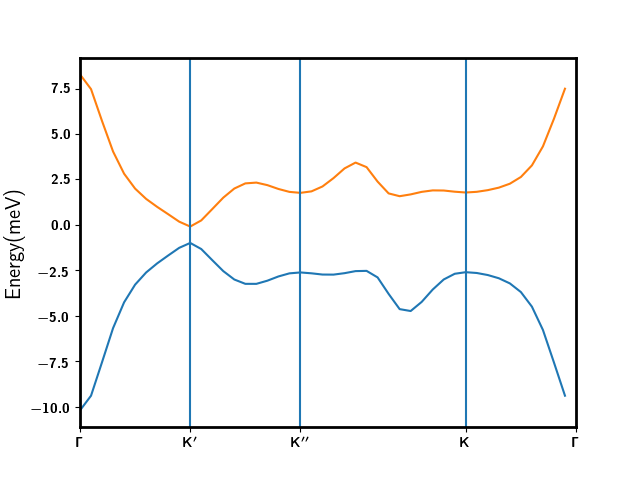}}
  
\caption{Dispersion for (a)Only $m_z$, (b)Case 1 and (c) Case 2. $\theta_G = 1.15^\circ$.}
\label{app:fig:dispersion}
\end{figure}

\begin{figure}[ht]
  \subfloat[]{
  \includegraphics[width = .32\textwidth]{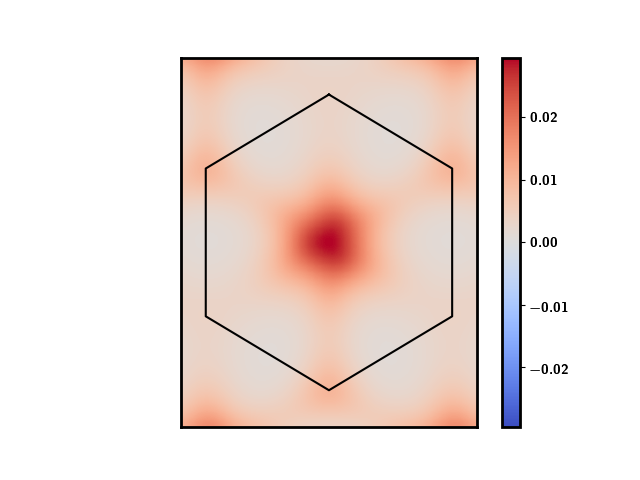}}
\subfloat[]{
  \includegraphics[width = .32\textwidth]{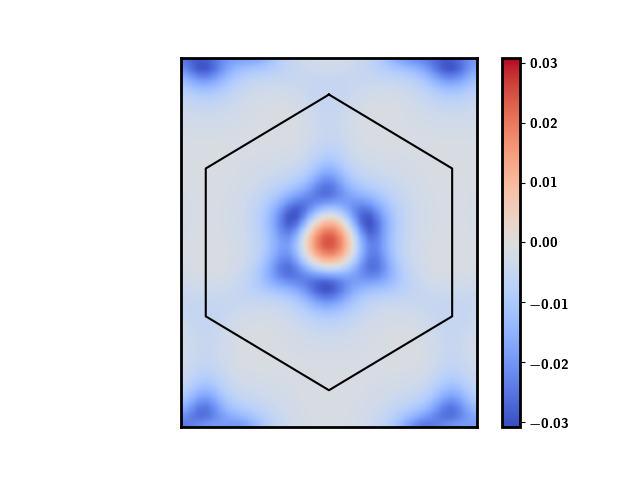}}
\subfloat[]{
  \includegraphics[width = .32\textwidth]{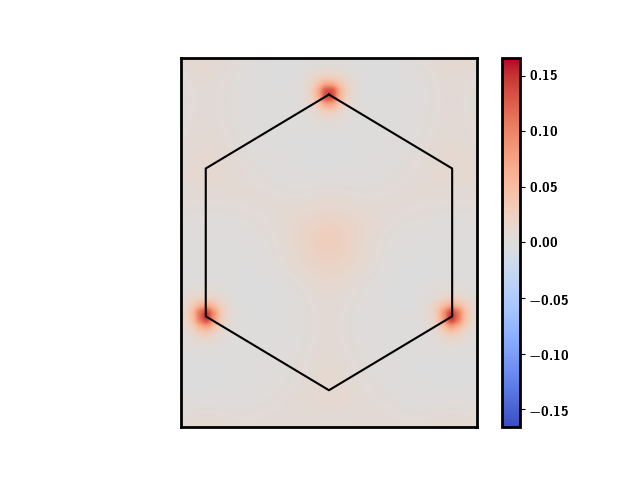}}
 
  \caption{Berry curvature distribution of the valence band for (a)Only $m_z$, (b)Case 1 and (c) Case 2. $\theta_G = 1.15^\circ$.}
\label{app:fig:berry}
\end{figure}

We plot the dispersion in Fig.\ref{app:fig:dispersion} and berry curvature distribution of the valence band in Fig.\ref{app:fig:berry}.

\subsection{Incommensurate Alignment}
\begin{figure}[ht]
  \subfloat[]{
  \includegraphics[width = .32\textwidth]{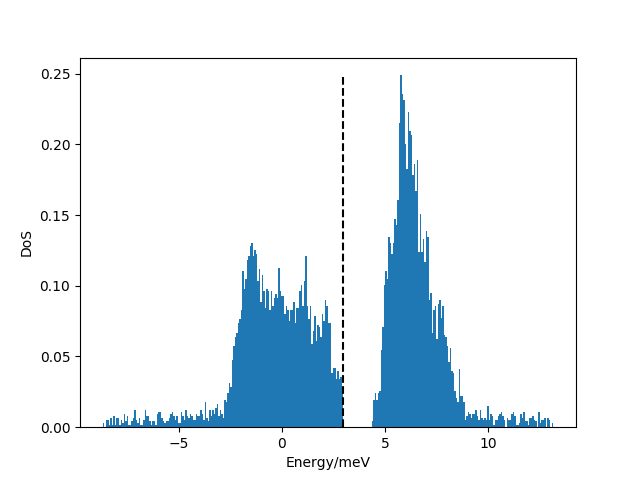}}
\subfloat[]{
  \includegraphics[width = .32\textwidth]{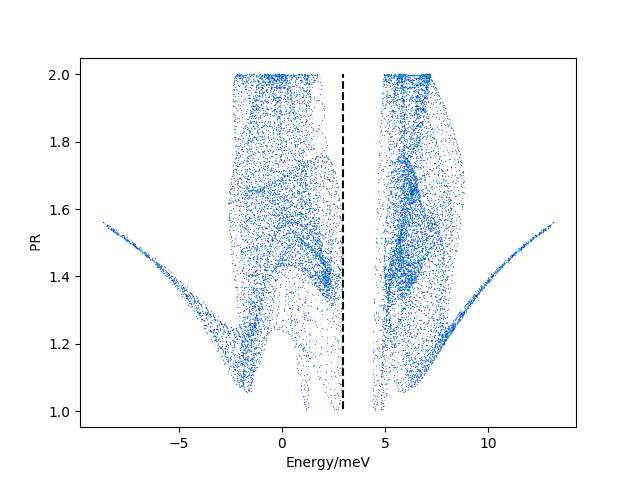}}
\subfloat[]{
  \includegraphics[width = .32\textwidth]{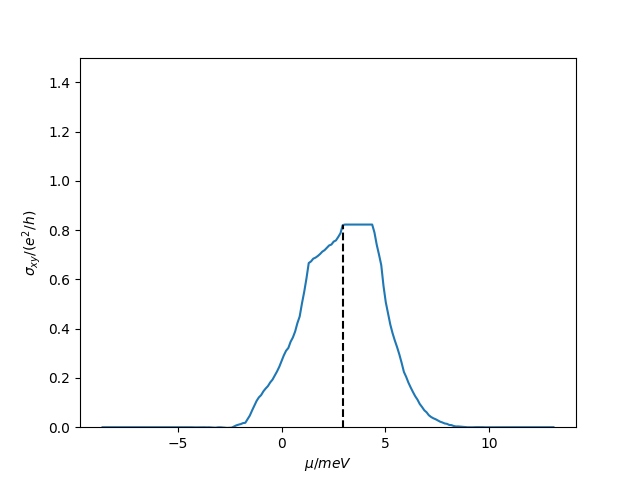}}

  \caption{Density of states, PR and $\sigma_{xy}$ for $\theta_G =1.15^\circ$ and $\theta_{BN} = -0.6^\circ$}
\label{app:fig:dosIPRsigmaxy}
\end{figure}

We study the density of states and PR for $\theta_G = 1.15^\circ$ and $\theta_{BN} = -0.6^\circ$, which is close to perfect alignment. Indeed we find a clear gap near charge neutrality and there is no indication of localization from PR. (See Fig.\ref{app:fig:dosIPRsigmaxy})

\section{Minimal coupling in a topological band}
\label{appendixB}

The gauge transformation operator can be written as,
\begin{equation}
    U = e^{i \int_{\vec{q}} \theta(\vec{q},\tau)\rho(\vec{q})},
\end{equation}
where $\rho(\vec{q})$ is the projected density operator in momentum space,
\begin{equation}
    \rho(\vec{q}) = \int \frac{d^2 k}{(2\pi)^2} c_{\vec{k}-\vec{q}}^\dag c_{\vec{k}} \lambda(\vec{k},\vec{q}).
\end{equation}
$\lambda(\vec{k},\vec{q}) = \langle u_{\vec{k}}| u_{\vec{k}-\vec{q}}\rangle$ is the form factor (note that the definition is different from the main text). And $\theta(\vec{q},\tau)$ is the Fourier transform of a real function $\theta(\vec{x},\tau)$, so $\theta(\vec{q},\tau)^* = \theta(-\vec{q},\tau)$. 

For an infinitesimal gauge transformation, $U \approx 1+ i \int_{\vec{q}} \theta(\vec{q},\tau) \rho(\vec{q})$ and an operator $\hat{O}\rightarrow U^\dag \hat{O} U$ under gauge transformation. Thus we have,
\begin{equation}
    \begin{split}
        U^\dag c_{\vec{k}} U &\approx c_{\vec{k}} +i \int_{\vec{q}} \theta(\vec{q},\tau) \lambda(\vec{k}+\vec{q},\vec{q})c_{\vec{k}+\vec{q}}\\
        U^\dag c_{\vec{k}}^\dag U &\approx c_{\vec{k}}^\dag -i \int_{\vec{q}} \theta(\vec{q},\tau) \lambda(\vec{k},\vec{q})c_{\vec{k}-\vec{q}}^\dag
    \end{split}
\end{equation}
Consider an action in Euclidean signature that contains three terms, $\mathcal{S}_0 =\int d^2 k \int d\tau  c_{\vec{k}}^\dag \partial_{\tau} c_{\vec{k}} - \epsilon_{\vec{k}} c_{\vec{k}}^\dag c_{\vec{k}} - \sum_{\vec{Q}} V_{\vec{Q}} \rho_{\vec{Q}}$, where $\vec{Q}$ does not need to be commensurate with the reciprocal lattice vector. Performing a gauge transformation to these terms, we have,
\begin{equation}
\begin{split}
       \delta (\int_{\vec{k}} c_{\vec{k}}^\dag \partial_{\tau} c_{\vec{k}}) &\approx -i \int_{\vec{k},\vec{q}} \theta(\vec{q},\tau) \lambda(\vec{k},\vec{q})c_{\vec{k}-\vec{q}}^\dag \partial_{\tau} c_{\vec{k}} +i \int_{\vec{k},\vec{q}}c_{\vec{k}}^\dag \partial_{\tau}[ \theta(\vec{q},\tau) \lambda(\vec{k}+\vec{q},\vec{q})c_{\vec{k}+\vec{q}}] \\
       &=i \int_{\vec{k},\vec{q}} [\partial_\tau \theta(\vec{q},\tau)] \lambda(\vec{k},\vec{q}) c_{\vec{k}-\vec{q}}^\dag c_{\vec{k}} \\
        \delta (\int_{\vec{k}}\epsilon_{\vec{k}}c_{\vec{k}}^\dag  c_{\vec{k}}) &\approx -i \int_{\vec{k},\vec{q}} \theta(\vec{q},\tau) \lambda(\vec{k},\vec{q})c_{\vec{k}-\vec{q}}^\dag \epsilon_{\vec{k}} c_{\vec{k}} +i \int_{\vec{k},\vec{q}}c_{\vec{k}}^\dag \epsilon_{\vec{k}} \theta(\vec{q},\tau) \lambda(\vec{k}+\vec{q},\vec{q})c_{\vec{k}+\vec{q}} \\
        &= i \int_{\vec{k},\vec{q}} \theta(\vec{q},\tau) \lambda(\vec{k},\vec{q})c_{\vec{k}-\vec{q}}^\dag c_{\vec{k}}(\epsilon_{\vec{k}-\vec{q}}-\epsilon_{\vec{k}})\\
        \delta(V_{\vec{Q}} \rho_{\vec{Q}})&\approx -i V_{\vec{Q}}\int_{\vec{k},\vec{q}} \lambda(\vec{k},\vec{Q}) \theta(\vec{q},\tau)\lambda(\vec{k}-\vec{Q},\vec{q})c_{\vec{k}-\vec{q}-\vec{Q}}^\dag c_{\vec{k}}+i V_{\vec{Q}}\int_{\vec{k},\vec{q}} \lambda(\vec{k},\vec{Q}) c_{\vec{k}-\vec{Q}}^\dag \theta(\vec{q},\tau)\lambda(\vec{k}+\vec{q},\vec{q})c_{\vec{k}+\vec{q}}\\
        &= i V_{\vec{Q}}\int_{\vec{k},\vec{q}} \theta(\vec{q},\tau) c_{\vec{k}-\vec{q}-\vec{Q}}^\dag c_{\vec{k}}[\lambda(\vec{k}-\vec{q},\vec{Q})\lambda(\vec{k},\vec{q})- \lambda(\vec{k},\vec{Q})\lambda(\vec{k}-\vec{Q},\vec{q})]
\end{split}
\end{equation}
If we further consider long wave-length gauge transformation, we only need to take small $\vec{q}$ in $\theta(\vec{q},\tau)$ into account. Note that $\partial_{\vec{k}} \epsilon_{\vec{k}} = \vec{v}_{\vec{k}}$ and 
\begin{equation}
    \lambda(\vec{k}-\vec{q},\vec{Q})\lambda(\vec{k},\vec{q})- \lambda(\vec{k},\vec{Q})\lambda(\vec{k}-\vec{Q},\vec{q}) = -\vec{q}\cdot [\vec{\partial}_{\vec{k}}\lambda(\vec{k},\vec{Q}) + i \lambda(\vec{k},\vec{Q})(\vec{\mathcal{A}}_{\vec{k}}-\vec{\mathcal{A}}_{\vec{k}-\vec{Q}})]+o(q^2),
\end{equation}
where $\vec{\mathcal{A}}_{\vec{k}} = -i \langle u_{\vec{k}}|\partial_{\vec{k}}u_{\vec{k}}\rangle$ is the Berry connection in momentum space.

Thus, the change in action $\mathcal{S}_0$ can be written as,
\begin{equation}
\begin{split}
        \delta\mathcal{S}_0 = &\int d\tau \int_{\vec{k},\vec{q}} (i \partial_{\tau} \theta(\vec{q},\tau)+i\vec{q}\cdot \vec{v}_{\vec{k}} \theta(\vec{q},\tau))\lambda(\vec{k},\vec{q})c^\dag_{\vec{k}-\vec{q}} c_{\vec{k}}\\
        &+ \sum_{\vec{Q}} iV_{\vec{Q}}\int_{\vec{k},\vec{q}} \theta(\vec{q},\tau) c_{\vec{k}-\vec{q}-\vec{Q}}^\dag c_{\vec{k}}\vec{q}\cdot [\vec{\partial}_{\vec{k}}\lambda(\vec{k},\vec{Q}+\vec{q}) + i \lambda(\vec{k},\vec{Q}+\vec{q})(\vec{\mathcal{A}}_{\vec{k}}-\vec{\mathcal{A}}_{\vec{k}-\vec{Q}-\vec{q}})]+ o(q^2)
\end{split}
\end{equation}

Now let us consider the electromagnetic potential $A(\vec{q},\tau) = (A_0,\vec{A})$. Note that the projection to the topological band should only affect the gauge transformation of the projected degrees of freedom and the gauge transformation of the electromagnetic potential should remain the same as before the projection. Thus under a gauge transformation, we have,
\begin{equation}
\begin{split}
        A_0(\vec{q},\tau) \rightarrow A_0(\vec{q},\tau) - \partial_\tau \theta(\vec{q},\tau)\\
        \vec{A}(\vec{q},\tau) \rightarrow \vec{A}(\vec{q},\tau) + i\vec{q}~\theta(\vec{q},\tau).
\end{split}
\end{equation}
The goal is to construct terms involving the electromagnetic potential $A(\vec{q},\tau)$ such that $\delta \mathcal{S}_0$ can be cancelled by the gauge transformation of the $A(\vec{q},\tau)$ field. As a first attempt, we consider the following action,
\begin{equation}
\begin{split}
       \mathcal{S}_1 = &\int d\tau \int_{\vec{k},\vec{q}}  (i A_0(\vec{q},\tau)- \vec{A}(\vec{q},\tau)\cdot \vec{v}_{\vec{k}}) \lambda(\vec{k},\vec{q}) c^\dag_{\vec{k}-\vec{q}} c_{\vec{k}}\\
    -&\int d\tau \int_{\vec{k},\vec{q}} \vec{A}(\vec{q},\tau)\cdot\sum_{\vec{Q}} V_{\vec{Q}} c_{\vec{k}-\vec{q}-\vec{Q}}^\dag c_{\vec{k}} [\vec{\partial}_{\vec{k}}\lambda(\vec{k},\vec{Q}+\vec{q}) + i \lambda(\vec{k},\vec{Q}+\vec{q})(\vec{\mathcal{A}}_{\vec{k}}-\vec{\mathcal{A}}_{\vec{k}-\vec{Q}-\vec{q}})].
\end{split}
\end{equation}
Note that $\delta\mathcal{S}_0$ is cancelled  by terms in $\delta\mathcal{S}_1$ but there are other terms in $\delta\mathcal{S}_1$ so we have,
\begin{equation}
    \begin{split}
        \delta \mathcal{S}_1 = -\delta \mathcal{S}_0 + \delta\mathcal{S}_1'
    \end{split}
\end{equation}
and 
\begin{equation}
\begin{split}
    &-i\delta\mathcal{S}_1' = \int d\tau \int_{\vec{k},\vec{q},\vec{q}'}(i A_0(\vec{q},\tau)- \vec{A}(\vec{q},\tau)\cdot \vec{v}_{\vec{k}})\lambda(\vec{k},\vec{q})[-c_{\vec{k}-\vec{q}-\vec{q}'}^\dag c_{\vec{k}}\theta(\vec{q}',\tau)\lambda(\vec{k}-\vec{q},\vec{q}')+c_{\vec{k}-\vec{q}}^\dag c_{\vec{k}+\vec{q}'}\theta(\vec{q}',\tau)\lambda(\vec{k}+\vec{q}',\vec{q}')]\\
   &-\int d\tau \int_{\vec{k},\vec{q},\vec{q}'} \vec{A}(\vec{q},\tau)\cdot\sum_{\vec{Q}} V_{\vec{Q}}[\vec{\partial}_{\vec{k}}\lambda(\vec{k},\vec{Q}+\vec{q}) + i \lambda(\vec{k},\vec{Q}+\vec{q})(\vec{\mathcal{A}}_{\vec{k}}-\vec{\mathcal{A}}_{\vec{k}-\vec{Q}-\vec{q}})]\\
   &\times[-c_{\vec{k}-\vec{q}-\vec{Q}-\vec{q}'}^\dag c_{\vec{k}}\theta(\vec{q}',\tau)\lambda(\vec{k}-\vec{q}-\vec{Q},\vec{q}')+c_{\vec{k}-\vec{q}-\vec{Q}}^\dag c_{\vec{k}+\vec{q}'}\theta(\vec{q}',\tau)\lambda(\vec{k}+\vec{q}',\vec{q}')]\\
    &=\int d\tau \int_{\vec{k},\vec{q},\vec{q}'} i A_0(\vec{q},\tau)c_{\vec{k}-\vec{q}-\vec{q}'}^\dag c_{\vec{k}} \theta(\vec{q}',\tau)[\lambda(\vec{k},\vec{q}')\lambda(\vec{k}-\vec{q}',\vec{q})-\lambda(\vec{k}-\vec{q},\vec{q}')\lambda(\vec{k},\vec{q})]\\
    &-\int d\tau \int_{\vec{k},\vec{q},\vec{q}'}  A_\mu(\vec{q},\tau)c_{\vec{k}-\vec{q}-\vec{q}'}^\dag c_{\vec{k}} \theta(\vec{q}',\tau)[v_{\vec{k}-\vec{q}'}^\mu \lambda(\vec{k},\vec{q}')\lambda(\vec{k}-\vec{q}',\vec{q})- v_{\vec{k}}^\mu \lambda(\vec{k}- \vec{q},\vec{q}')\lambda(\vec{k},\vec{q})]\\
    &-\int d\tau \int_{\vec{k},\vec{q},\vec{q}'} \vec{A}(\vec{q},\tau)\cdot\sum_{\vec{Q}} V_{\vec{Q}}c_{\vec{k}-\vec{q}-\vec{Q}-\vec{q}'}^\dag c_{\vec{k}}\theta(\vec{q}',\tau)\{[\vec{\partial}_{\vec{k}-\vec{q}'}\lambda(\vec{k}-\vec{q}',\vec{Q}+\vec{q}) + i \lambda(\vec{k}-\vec{q}',\vec{Q}+\vec{q})(\vec{\mathcal{A}}_{\vec{k}-\vec{q}'}-\vec{\mathcal{A}}_{\vec{k}-\vec{q}'-\vec{Q}-\vec{q}})]\lambda(\vec{k},\vec{q}')\\
    &-[\vec{\partial}_{\vec{k}}\lambda(\vec{k},\vec{Q}+\vec{q}) + i \lambda(\vec{k},\vec{Q}+\vec{q})(\vec{\mathcal{A}}_{\vec{k}}-\vec{\mathcal{A}}_{\vec{k}-\vec{Q}-\vec{q}})]\lambda(\vec{k}-\vec{q}-\vec{Q},\vec{q}')\}\\
    &\approx -\int d\tau \int_{\vec{k},\vec{q},\vec{q}'} (i A_0(\vec{q},\tau)-\vec{A}(\vec{q},\tau)\cdot \vec{v}_{\vec{k}}) c_{\vec{k}-\vec{q}-\vec{q}'}^\dag c_{\vec{k}} \theta(\vec{q}',\tau) \vec{q}' \cdot[\vec{\partial}_{\vec{k}}\lambda(\vec{k},\vec{q})+i\lambda(\vec{k},\vec{q})(\vec{\mathcal{A}}_{\vec{k}}-\vec{\mathcal{A}}_{\vec{k}-\vec{q}})]\\
    &+\int d\tau \int_{\vec{k},\vec{q},\vec{q}'}  A_\mu(\vec{q},\tau)c_{\vec{k}-\vec{q}-\vec{q}'}^\dag c_{\vec{k}} \theta(\vec{q}',\tau) \lambda(\vec{k},\vec{q}+\vec{q}')\vec{q}' \cdot \vec{\partial}_{\vec{k}}v^\mu_{\vec{k}}\\
    &-\int d\tau \int_{\vec{k},\vec{q},\vec{q}'} \vec{A}(\vec{q},\tau)\cdot\sum_{\vec{Q}} V_{\vec{Q}}c_{\vec{k}-\vec{q}-\vec{Q}-\vec{q}'}^\dag c_{\vec{k}}\theta(\vec{q}',\tau)\{-(\vec{q}'\cdot \vec{\partial}_{\vec{k}})\vec{\partial}_{\vec{k}}\lambda(\vec{k},\vec{Q}+\vec{q})  -i\vec{q}'\cdot  \vec{\partial}_{\vec{k}}[\lambda(\vec{k},\vec{Q}+\vec{q})(\vec{\mathcal{A}}_{\vec{k}}-\vec{\mathcal{A}}_{\vec{k}-\vec{Q}-\vec{q}})]\\
    &+[\vec{\partial}_{\vec{k}}\lambda(\vec{k},\vec{Q}+\vec{q}) + i \lambda(\vec{k},\vec{Q}+\vec{q})(\vec{\mathcal{A}}_{\vec{k}}-\vec{\mathcal{A}}_{\vec{k}-\vec{Q}-\vec{q}})]i\vec{q}' \cdot (\vec{\mathcal{A}}_{\vec{k}-\vec{q}-\vec{Q}}-\vec{\mathcal{A}}_{\vec{k}})\}+o((q')^2)\\
    &\approx i\int d\tau \int_{\vec{k},\vec{q},\vec{q}'} (i A_0(\vec{q},\tau)-\vec{A}(\vec{q},\tau)\cdot \vec{v}_{\vec{k}}) c_{\vec{k}-\vec{q}-\vec{q}'}^\dag c_{\vec{k}}\lambda(\vec{k},\vec{q}+\vec{q}') \theta(\vec{q}',\tau) \vec{q}'\times \vec{q}\cdot \hat{z} \mathcal{B}_{\vec{k}} \\
    &+\int d\tau \int_{\vec{k},\vec{q},\vec{q}'}  A_\mu(\vec{q},\tau)c_{\vec{k}-\vec{q}-\vec{q}'}^\dag c_{\vec{k}} \theta(\vec{q}',\tau) \lambda(\vec{k},\vec{q}+\vec{q}')\vec{q}' \cdot \vec{\partial}_{\vec{k}}v^\mu_{\vec{k}}\\
    &-\int d\tau \int_{\vec{k},\vec{q},\vec{q}'} \sum_{\vec{Q}} V_{\vec{Q}}c_{\vec{k}-\vec{Q}-\vec{q}-\vec{q}'}^\dag c_{\vec{k}}\theta(\vec{q}',\tau)
    [\vec{q}'\cdot(-i\vec{\partial}_{\vec{k}} +  \vec{\mathcal{A}}_{\vec{k}}-\vec{\mathcal{A}}_{\vec{k}-\vec{Q}-\vec{q}-\vec{q}'})]\\&\times[\vec{A}(\vec{q},\tau)\cdot(-i\vec{\partial}_{\vec{k}} +  \vec{\mathcal{A}}_{\vec{k}}-\vec{\mathcal{A}}_{\vec{k}-\vec{Q}-\vec{q}-\vec{q}'})]\lambda(\vec{k},\vec{Q}+\vec{q}+\vec{q}')+o((q')^2)+o(q^2),
\end{split}
\label{eq:deltaS1'}
\end{equation}
where $\mathcal{B}_{\vec{k}} = \partial_{k_x}\mathcal{A}_{\vec{k},y}-\partial_{k_y}\mathcal{A}_{\vec{k},x}$ is the Berry curvature.

Now, let us further consider the possible terms that cancel $\delta\mathcal{S}'_1$. Suppose there is $\mathcal{S}_2$, and
\begin{equation}
    \begin{split}
    -i\mathcal{S}_2 =& -\int d\tau \int_{\vec{k},\vec{q},\vec{q}'} i A_0(\vec{q},\tau) (\vec{A}(\vec{q}',\tau) \times \vec{q})\cdot\hat{z} c_{\vec{k}-\vec{q}-\vec{q}'}^\dag c_{\vec{k}} \lambda(\vec{k},\vec{q}+\vec{q}') \mathcal{B}_{\vec{k}}\\
    +&\int d\tau \int_{\vec{k},\vec{q},\vec{q}'} (1/2  (\partial_{\tau}\vec{A}(\vec{q},\tau) \times \vec{A}(\vec{q}',\tau)) )\cdot\hat{z} c_{\vec{k}-\vec{q}-\vec{q}'}^\dag c_{\vec{k}} \lambda(\vec{k},\vec{q}+\vec{q}')\mathcal{B}_{\vec{k}}\\
    +&\int d\tau \int_{\vec{k},\vec{q},\vec{q}'}  A_\mu(\vec{q},\tau)c_{\vec{k}-\vec{q}-\vec{q}'}^\dag c_{\vec{k}}  \lambda(\vec{k},\vec{q}+\vec{q}')i A_\nu(\vec{q}',\tau) \partial_{k_\nu}v^\mu_{\vec{k}}/2+ (\vec{A}(\vec{q},\tau)\cdot \vec{v}_{\vec{k}})  (\vec{A}(\vec{q}',\tau) \times \vec{q})\cdot\hat{z} c_{\vec{k}-\vec{q}-\vec{q}'}^\dag c_{\vec{k}} \lambda(\vec{k},\vec{q}+\vec{q}') \mathcal{B}_{\vec{k}}\\
    +&\frac{-i}{2}\int d\tau \int_{\vec{k},\vec{q},\vec{q}'} \sum_{\vec{Q}} V_{\vec{Q}}c_{\vec{k}-\vec{Q}-\vec{q}-\vec{q}'}^\dag c_{\vec{k}}\{\vec{A}(\vec{q}',\tau)\cdot[ -i\vec{\partial}_{\vec{k}}+\vec{\mathcal{A}}_{\vec{k}}-\vec{\mathcal{A}}_{\vec{k}-\vec{Q}-\vec{q}-\vec{q}'}]\}\\
    &\times\{\vec{A}(\vec{q},\tau)\cdot[ -i\vec{\partial}_{\vec{k}}+\vec{\mathcal{A}}_{\vec{k}}-\vec{\mathcal{A}}_{\vec{k}-\vec{Q}-\vec{q}-\vec{q}'}]\}\lambda(\vec{k},\vec{Q}+\vec{q}+\vec{q}').
    \end{split}
    \label{eq:S2}
\end{equation}
One can verify that $\delta \mathcal{S}_2$ cancels $\delta \mathcal{S}_1'$ for small $\vec{q}$ and $\vec{q}'$.

The first two terms in Eq.\ref{eq:S2} can be combined to a Chern-Simons(CS) term. To see this, define $\phi_{\mathcal{B}}(\vec{q}) = \int_{\vec{k}} c_{\vec{k}-\vec{q}}^\dag c_{\vec{k}}\lambda(\vec{k},\vec{q}) \mathcal{B}_{\vec{k}}$ and in real space, the first two terms in Eq.\ref{eq:S2} reduce to,
\begin{equation}
    \frac{1}{2}\int d\tau \int d^2 x \phi_{\mathcal{B}}(-\vec{x}) A(\vec{x},\tau) dA(\vec{x},\tau) ,
\end{equation}
where $\phi_{\mathcal{B}}(\vec{x})$ is the Fourier transformation of $\phi_{\mathcal{B}}(\vec{q})$.

As a sanity check, consider a Chern insulator. The ground state expectation value of $\phi_{\mathcal{B}}(\vec{x})$ is $\langle \phi_{\mathcal{B}}(\vec{x})\rangle_{G.S.} = \frac{C}{2\pi}$, where $C$ is the Chern number. This gives the correct quantized coefficient for the CS term.

Thus, at long wave-length, the action for a topological band that is minimally coupled to gauge field is,
\begin{equation}
    \mathcal{S}[A] = \mathcal{S}_0 + \mathcal{S}_1[A] +\mathcal{S}_2[A]
\end{equation}
The current density operator is $\vec{J} =  \frac{\delta \mathcal{L}}{\delta \vec{A}}$. Let us consider applying an external static electric field and choose a gauge such that $\vec{E} = -\vec{\nabla} A_0(\vec{x},t)$ and $\vec{A}(\vec{x},t)=0$. The current  density operator around $\vec{q}=0$ is thus,
\begin{equation}
\begin{split}
    J^\mu(-\vec{q}) &\approx- \int_{\vec{k}} v_{\vec{k}}^\mu c_{\vec{k}-\vec{q}}^\dag c_{\vec{k}} \lambda(\vec{k},\vec{q}) - \phi_{\mathcal{B}}(\vec{q}=0)\epsilon_{\mu\nu} E^\nu\\
    &-i\sum_{\vec{Q}} V_{\vec{Q}}\int_{\vec{k}} c_{\vec{k}-\vec{Q}-\vec{q}}^\dag c_{\vec{k}} [-i\partial^\mu_{\vec{k}} + (\mathcal{A}_{\vec{k}}^\mu - \mathcal{A}_{\vec{k}-\vec{Q}-\vec{q}}^\mu)]\lambda(\vec{k},\vec{Q}+\vec{q})
\end{split}
\end{equation}
For a flat topological band, $\vec{v}_{\vec{k}}=0$ so the total current $\vec{\mathcal{J}}$ is,
\begin{equation}
    \begin{split}
        \mathcal{J}^\mu  = J^\mu(\vec{q}=0)=-\epsilon^{\mu\nu} \sum_{\vec{k}} c_{\vec{k}}^\dag c_{\vec{k}} \mathcal{B}_{\vec{k}} E_\nu-i\sum_{\vec{Q}} V_{\vec{Q}}\sum_{\vec{k}} c_{\vec{k}-\vec{Q}}^\dag c_{\vec{k}} [-i\partial^\mu_{\vec{k}} + (\mathcal{A}_{\vec{k}}^\mu - \mathcal{A}_{\vec{k}-\vec{Q}}^\mu)]\lambda(\vec{k},\vec{Q})
    \end{split}
    \label{eq:total_current}
\end{equation}
In the main text, we take $\lambda(\vec{k},\vec{Q}) = e^{i \int_{\vec{k}}^{\vec{k}-\vec{Q}}\vec{\mathcal{A}}}$ so we have,
\begin{equation}
    [-i\partial^\mu_{\vec{k}} + (\mathcal{A}_{\vec{k}}^\mu - \mathcal{A}_{\vec{k}-\vec{Q}}^\mu)]\lambda(\vec{k},\vec{Q}) = \lambda(\vec{k},\vec{Q})\int_{\vec{k}}^{\vec{k}-\vec{Q}} \mathcal{B}(\vec{k}') \epsilon^{\mu\nu} dk'_{\nu}.
\end{equation}
If the Berry curvature is uniform, the above expression will reduce to $-\epsilon^{\mu\nu}Q_{\nu}\lambda(\vec{k},\vec{Q})\mathcal{B}$ so we find that Eq.\ref{eq:total_current} agrees with Eq.\ref{eq:current} in the main text.

To conclude, in the derivation of Eq.\ref{eq:total_current} we don't assume any specific form of $\lambda(\vec{k},\vec{Q})$ so the expression of the current operator can be used in any topological band with non-trivial Berry curvature.

\section{Bloch wave function in LLL and form factors}
\label{wavefn}
In a Landau gauge $\vec{A} = (-B y,0)$, the magnetic translation operator are,
\begin{equation}
    \begin{cases}
        T_x &= e^{i P_x}\\
        T_y &= e^{i (P_y +B x)}.
    \end{cases}
\end{equation}
One can verify that $[T_x,T_y]=0$ and $T_x$, $T_y$ commute with the kinetic momenta $\vec{P}-\vec{A}$, since the magnetic flux $\Phi = B = 2\pi$, where we set the lattice constant to 1.

The eigenfunction $\psi(\vec{x})$ of $T_{x,y}$ can be labeled by the momenta $(k_x,k_y)$ such that the eigenvalues are $e^{i k_{x,y}}$. In order to construct such eigenfunctions, we first examine how $T_{x,y}$ act on a wave function in LLL, that is $\phi_{k_x}(\vec{x}) = e^{i k_x x} \Psi_0(y+\frac{k_x}{B})$, where $\Psi_0(y)=(\frac{B}{\pi})^{\frac{1}{4}}e^{-\frac{B y^2}{2}}$. It is readily seen that $\phi_{k_x}(\vec{x})$ is an eigenfunction of $T_x$ and $T_y \phi_{k_x}(\vec{x}) = \phi_{k_x+B}(\vec{x})$. Thus the eigenfunction of $T_{x,y}$ can be written as,
\begin{equation}
    \psi_{\vec{k}}(\vec{x}) = 
\sum_{m=-\infty}^\infty e^{-i m k_y-i k_x k_y/B} \phi_{k_x+mB}(\vec{x})= \sum_{m=-\infty}^\infty e^{i (k_x x + mB x-mk_y-k_x k_y/B)} \Psi_0(y+\frac{k_x}{B}+m),
\end{equation}
where $m$ is an integer. Note that we choose a gauge in $\vec{k}$ space such that $\psi_{(k_x,k_y+2\pi)}(\vec{x}) = e^{-i k_x}\psi_{(k_x,k_y)}(\vec{x})$ and $\psi_{(k_x+2\pi,k_y)}(\vec{x}) = \psi_{(k_x,k_y)}(\vec{x})$.

For an opposite magnetic field, we can choose a Landau gauge such that $\tilde{\vec{A}} = (By,0)$. Then the wavefunction in the LLL can be written as $\tilde{\phi}_{k_x}(\vec{x}) = e^{i k_x x} \Psi_0(y-\frac{k_x}{B})$. And the magnetic translation operators are $\tilde{T}_x = T_x$, $\tilde{T}_y = e^{i(P_y-Bx)}$ so the corresponding eigenfunction can be written as,
\begin{equation}
    \tilde{\psi}_{\vec{k}}(\vec{x})=\sum_{m=-\infty}^{\infty} e^{i m k_y+i k_x k_y/B} \tilde{\phi}_{k_x+mB}(\vec{x})= \sum_{m=-\infty}^\infty e^{i (k_x x + mB x+mk_y+k_x k_y/B)} \Psi_0(y-\frac{k_x}{B}-m).
\end{equation}

Now let us consider various form factors,
\begin{equation}
\begin{split}
        &\lambda_{\pm\pm} (\vec{k},\vec{k}+\vec{Q}) = \sum_{m,m'}\int dx \int dy e^{\mp i (k_x Q_y+Q_x k_y + Q_x Q_y)/B-i(mB-m'B)x\mp im'(k_y+Q_y)\pm im k_y-iQ_y y}\\&\times \Psi_0(y\pm \frac{k_x+mB}{B})\Psi_0(y\pm \frac{k_x+m'B+Q_x}{B})\\
       &=\sum_m \int dy e^{\mp i (k_x Q_y+Q_x k_y + Q_x Q_y)/B \mp i m Q_y-iQ_y y}\Psi_0(y\pm \frac{k_x+mB}{B})\Psi_0(y\pm \frac{k_x+mB+Q_x}{B})\\
       &=\sum_m \int dy e^{\mp i (k_x Q_y+Q_x k_y + Q_x Q_y)/B \mp i m Q_y} [\Psi_0(y\pm \frac{(k_x+mB)}{B}+\frac{\pm Q_x +i Q_y}{2B})]^2 e^{-B(\frac{Q_x}{2B})^2+B(\frac{\pm Q_x+iQ_y}{2B})^2} e^{\pm\frac{(k_x+mB)iQ_y}{B}}\\
       &= e^{\mp i \frac{k_y Q_x}{B}\mp \frac{i Q_x Q_y}{2B} -\frac{Q_x^2+Q_y^2}{4B}}.
\end{split}
\end{equation}
Now let us consider $\lambda_{+-}^0(\vec{k}) = \langle \psi_{\vec{k};+}| \psi_{\vec{k};-}\rangle$.
\begin{equation}
\begin{split}
        \lambda_{+-}^0(\vec{k}) &= \sum_{m}\int dy e^{i m 2 k_y+i 2 k_x k_y/B}\Psi_0(y+\frac{k_x+mB}{B})\Psi_0(y-\frac{k_x+mB}{B})\\
        &= \sum_{m}\int dy e^{im 2k_y+i 2 k_x k_y/B}\left[\Psi_0(y)\right]^2 e^{-B(\frac{k_x+mB}{B})^2}\\
        &=\sum_{m}e^{2 i \frac{k_y}{B}(k_x +m B) -\frac{(k_x+mB)^2}{B}}.
\end{split}
\end{equation}

\section{Inverse participation ratio of various systems}
\label{sec:IPR}
\subsection{IPR of single band models}
\begin{figure}[ht]
\subfloat[]{
  \includegraphics[width = .48\textwidth]{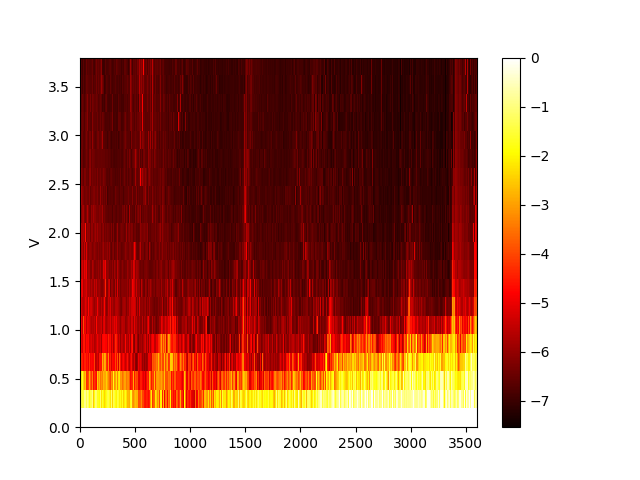}}
\subfloat[]{
  \includegraphics[width = .48\textwidth]{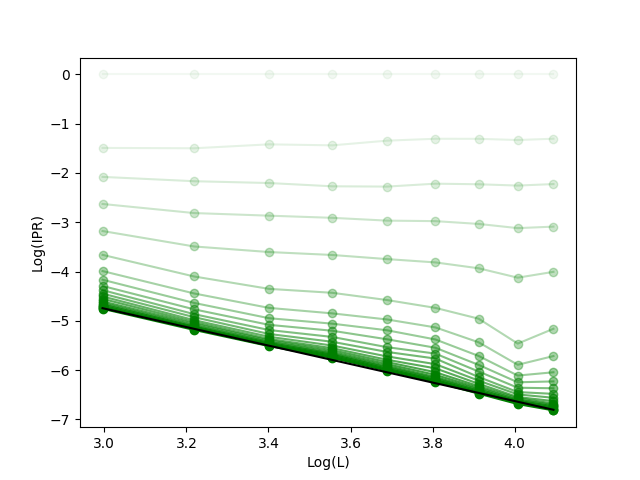}}
  
\subfloat[]{
  \includegraphics[width = .48\textwidth]{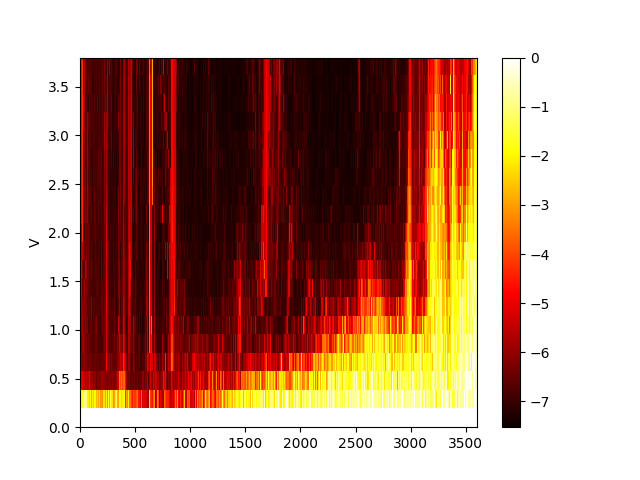}}
\subfloat[]{
  \includegraphics[width = .48\textwidth]{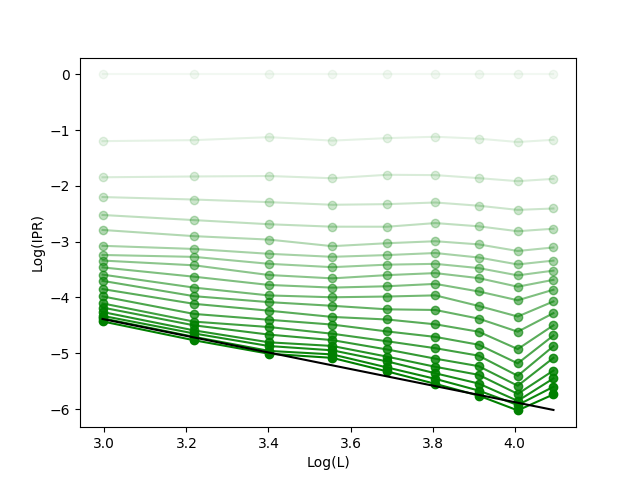}}

\caption{IPR for single band model with dispersion. (a),(b): trivial case. (c),(d): non-trivial case. In (a)(b), Log(IPR) is plotted, x-axis labels different eigenstates in ascending order of their energies and y-axis labels different QP potential V from 0 to 4. The system size is $60\times 60$. In (b)(d), green dots with different brightness connected by the same line label different QP potential V, from 0 to 4 (top to bottom), with 0.2 interval. The black line is the linear fitting for $V=4$. The linear system sizes are from 20 to 60. }
\label{app:fig:IPR_single_with_dispersion}
\end{figure}

We consider the Hamiltonian in Eq.\ref{eq:toymodel_1band} with $V_1=V_2=V_3=V$ and further include a dispersion term $H_{dis} = \sum_{\vec{k}} c_{\vec{k}}^\dag c_{\vec{k}} \epsilon_{\vec{k}}$, where we take $\epsilon_{\vec{k}} = t \sum_{j=1}^3 \cos{(\vec{k}\cdot \vec{b}_j)}$. $\vec{b}_j$'s are the lattice vectors, taken to be $\vec{b}_1 = \frac{4\pi Q}{\sqrt{3}} (0,-1)$ and $\vec{b}_{2,3} = \frac{4\pi Q}{\sqrt{3}} (\pm \frac{\sqrt{3}}{2},\frac{1}{2})$. We take the QP lattice to be parallel to the original lattice and $Q$ is an irrational number. We calculate the IPR for various QP potential strength $V$ and system sizes (Fig. \ref{app:fig:IPR_single_with_dispersion}) given $t=1$ and $Q=\sqrt{5}-1$. We compare the system with vanishing Berry curvature (trivial) and uniform Berry curvature given by a $C=1$ band (non-trivial). We find that in both cases, IPR has a stronger system size dependence with increasing QP potential, which indicates that there are more extended states in $\vec{k}$-space. Moreover, we fit the dependence of $Log(IPR)$ to the logarithm of linear system size $L$ and find a linear dependence, $Log(IPR) = z Log(L) + b$. And the trivial model has a slope $z\approx-1.88$, while for the non-trivial model, $z\approx-1.37$. The slope of the trivial model is closer to the ideal scaling $z=2$, which indicates that there are more localized states in real space in the trivial model than in the non-trivial model with large QP potential.

We further consider a non-uniform Berry curvature in $\vec{k}$-space with $B(\vec{k}) = B_0 + B_1\sum_{j=1}^3 \cos{(\vec{k}\cdot \vec{b}_j)}$, where $B_0$ is the uniform part as considered before. The $B_1$ term acts as a QP hopping term. We calculate the IPR with trivial and non-trivial $B_0$ and increasing $B_1$ (Fig.\ref{app:fig:IPR_single_with_dispersion}). For the trivial case, we find $z\approx-1.92$. For the non-trivial case, at small value of $B_1$, $z\approx -1.77$ and $z$ decreases with increasing $B_1$. $z \approx 1.12$ within $2\leq B_1\leq4$. The above results indicate that in the presence of non-uniform Berry curvature, there are more non-localized states (extended or critical) in real space in the non-trivial model compared to the trivial model.

\begin{figure}[h]
    \subfloat[]{
  \includegraphics[width = .48\textwidth]{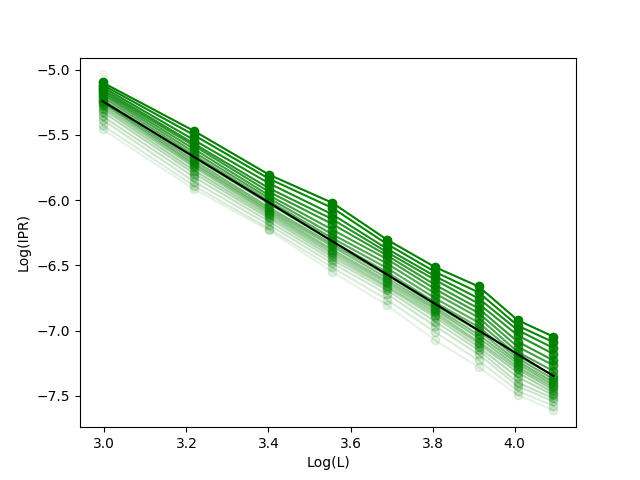}}
\subfloat[]{
  \includegraphics[width = .48\textwidth]{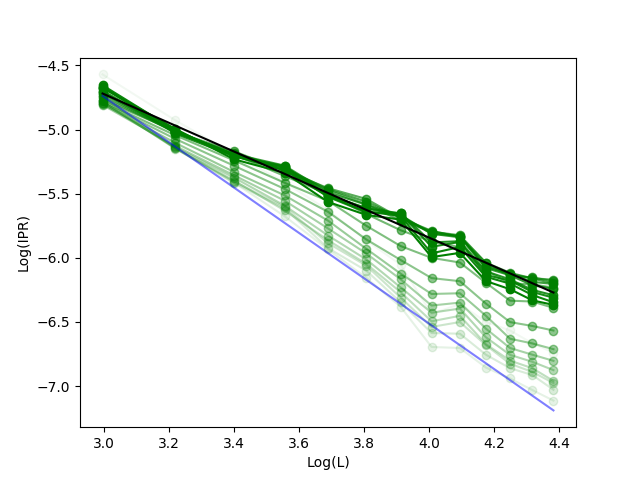}}
    \caption{IPR for single band model with non-uniform Berry curvature. (a): trivial($B_0=0$) and (b): non-trivial($B_0\approx 7.26$). Green dots with different brightness connected by the same line label different $B_1$, from 0 to 4 (top to bottom), with 0.2 interval. The black line in (a) is the linear fitting of the average IPR. The black line in (b) is the linear fitting of the average IPR for $B_1\geq2$ and the blue line in (b) is the linear fitting for $B_1=0.2$. }
    \label{app:fig:IPR_single_periodic_mag}
\end{figure}
\subsection{IPR of the two band model}
\begin{figure}[h]
  \subfloat[]{
  \includegraphics[width = .48\textwidth]{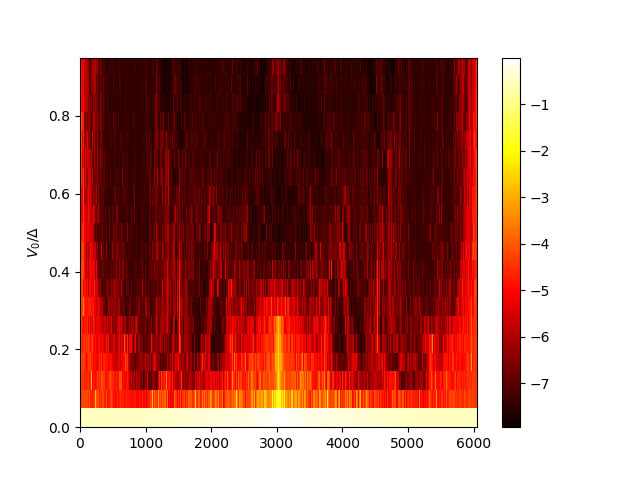}}
    \subfloat[]{
  \includegraphics[width = .48\textwidth]{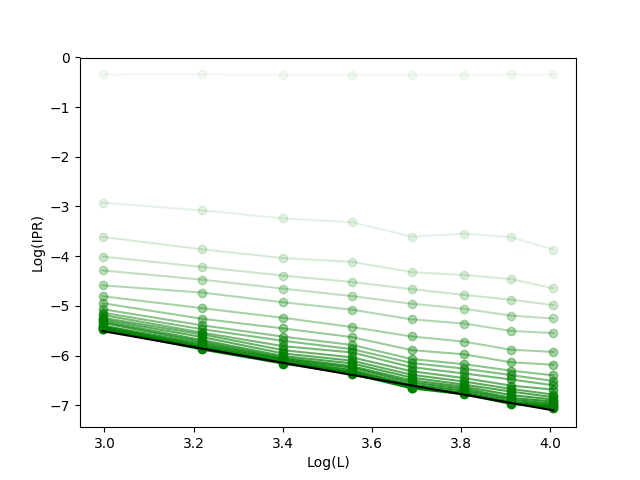}}

 \caption{IPR for the two band model. (a) Color plot of Log(IPR).X-axis labels different eigenvalues. (b) Dependence of Log(IPR) with linear system size. The green lines from bright to dark labels different $V_0$ from 0 to 1. The black line is the linear fitting to the Log(IPR) at $V_0=1$.}
 \label{app:fig:IPR_double_avg}
 \end{figure}

\begin{figure}[h]
  \subfloat[]{
  \includegraphics[width = .33\textwidth]{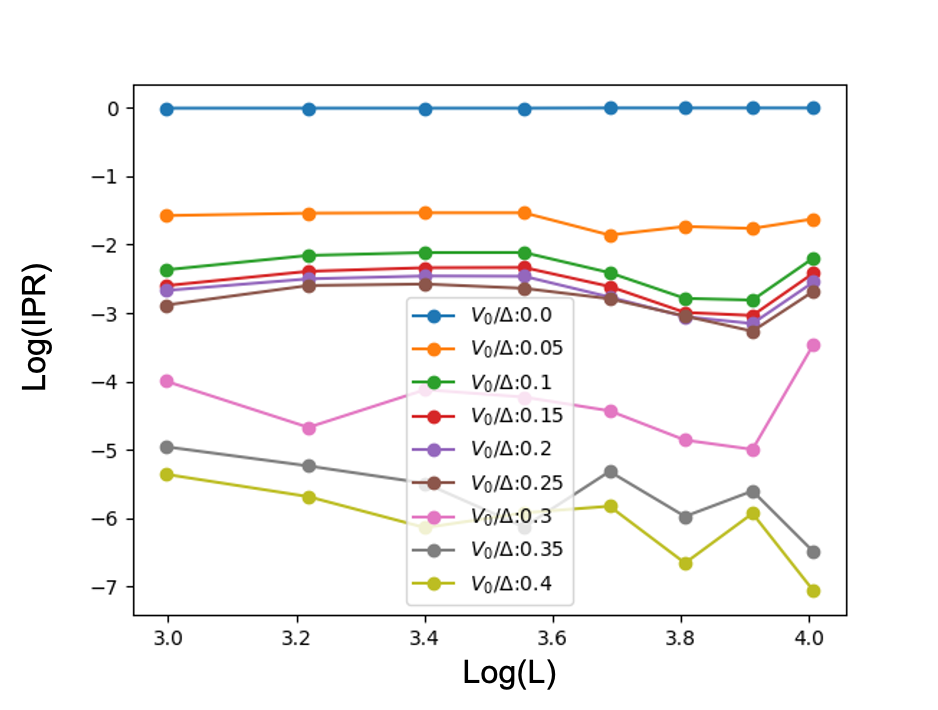}}
    \subfloat[]{
  \includegraphics[width = .33\textwidth]{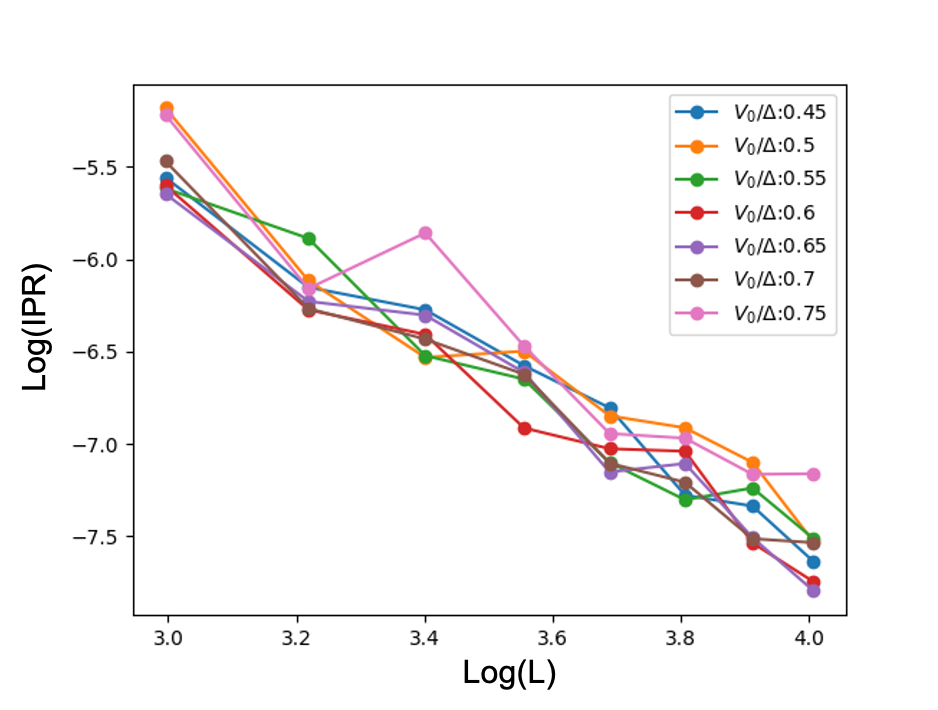}}
   \subfloat[]{
  \includegraphics[width = .33\textwidth]{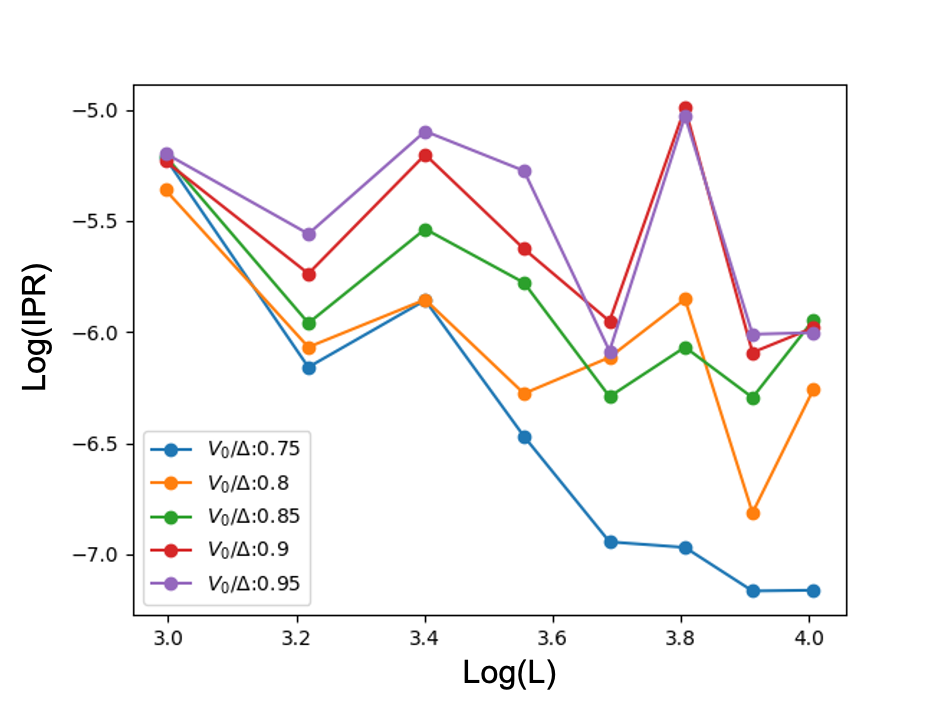}}

 \caption{IPR of the states that is in the middle of the spectrum. The x-axis is Log(L) and the y-axis is Log(IPR) for (a)small $V_0$, (b)intermediate $V_0$ and (c)large $V_0$.}
 \label{app:fig:IPR_double_sml}
 \end{figure}

We consider the Hamiltonian in Eq.\ref{eq:pmChern_Ham}. We let $V_1= \Delta=1$ and change $V_0$. The IPR's for various $V_0$ are shown in Fig.\ref{app:fig:IPR_double_avg}. We find that the slope of Log(IPR)-Log(L) curve goes from $z= 0$ to $z\approx -1.58$ when $V_0$ is increased from 0 to 1. At small $V_0$, there is a band gap in the middle of the spectrum. When $V_0/\Delta \gtrapprox 0.2$, the gap closes and the IPR values in the middle of the spectrum  get larger after the gap closes, which indicates that there are extended states in the middle of the spectrum and they get localized with increasing $V_0$. We study the IPR of the states that are in the middle of the spectrum (Fig.\ref{app:fig:IPR_double_sml}) and find that there is an intermediate regime of $V_0$ where the IPRs of the states can be fit to a power law dependence to the linear system size with a power $z\approx -1.97$. Thus, the states in the middle of the spectrum are indeed localized in real space.


\end{document}